\documentclass[review,12pt]{elsarticle}

\usepackage{amssymb}
\usepackage{amsthm}
\usepackage{framed} 
\usepackage{amsmath,color}
\usepackage{mathrsfs}
\usepackage{graphicx}
\usepackage{epstopdf}
\usepackage{float}
\usepackage{caption}
\usepackage{lineno}
\usepackage{subcaption}
\usepackage{bm}
\usepackage{bbm}
\usepackage{mathrsfs}
\usepackage{hyperref}
\usepackage{cleveref}
\usepackage{soul}
\usepackage{svg} 
\usepackage{accents}
\usepackage{color,soul} 
\usepackage{color} 
\usepackage{bm}
\usepackage{multirow} 
\usepackage[margin=2cm]{geometry}
\usepackage{adjustbox}
\usepackage{comment}
\usepackage{blindtext}
\usepackage{ragged2e}
\biboptions{sort&compress}
\soulregister\citep7 
\soulregister\citet7 
\soulregister\citealp7 
\newsavebox{\measurebox} 
\usepackage{titlesec} 
\usepackage[T1]{fontenc} 
\usepackage{lmodern} 
\input{glyphtounicode}  

\def\onedot{$\mathsurround0pt\ldotp$}
\def\cdddot#1{
  \mathbin{\vcenter{\baselineskip.67ex
    \hbox{\onedot}\hbox{\onedot}\hbox{\onedot}%
  }}%
}

\journal{International Journal of Mechanical Sciences}

\makeatletter
\def\@author#1{\g@addto@macro\elsauthors{\normalsize%
    \def\baselinestretch{1}%
    \upshape\authorsep#1\unskip\textsuperscript{%
      \ifx\@fnmark\@empty\else\unskip\sep\@fnmark\let\sep=,\fi
      \ifx\@corref\@empty\else\unskip\sep\@corref\let\sep=,\fi
      }%
    \def\authorsep{\unskip,\space}%
    \global\let\@fnmark\@empty
    \global\let\@corref\@empty  
    \global\let\sep\@empty}%
    \@eadauthor={#1}
}
\makeatother

\titleformat{\paragraph}
{\normalfont\normalsize\itshape}{\theparagraph}{1em}{}
\titlespacing*{\paragraph}
{0pt}{3.25ex plus 1ex minus .2ex}{1.5ex plus .2ex}

\graphicspath{{Figures/}}

\begin{document}

\begin{frontmatter}



\title{Coupled thermo-chemo-mechanical phase field-based modelling of hydrogen-assisted cracking in girth welds}


\author{Lucas Castro\fnref{Uniovi}}

\author{Yousef Navidtehrani \fnref{Uniovi}}

\author{Covadonga Betegón \fnref{Uniovi}}

\author{Emilio Mart\'{\i}nez-Pa\~neda\corref{cor1}\fnref{OXFORD}}
\ead{emilio.martinez-paneda@eng.ox.ac.uk}

\address[Uniovi]{Department of Construction and Manufacturing Engineering, University of Oviedo, Gij\'{o}n 33203, Spain}

\address[OXFORD]{Department of Engineering Science, University of Oxford, Oxford OX1 3PJ, UK}

\cortext[cor1]{Corresponding author.}

\begin{abstract}
A new computational framework is presented to predict the structural integrity of welds in hydrogen transmission pipelines. The framework combines: (i) a thermo-mechanical weld process model, and (ii) a coupled deformation-diffusion-fracture phase field-based model that accounts for plasticity and hydrogen trapping, considering multiple trap types, with stationary and evolving trap densities. This enables capturing, for the first time, the interplay between residual stresses, trap creation, hydrogen transport, and fracture. The computational framework is particularised and applied to the study of weld integrity in X80 pipeline steel. The focus is on girth welds, as they are more complex due to their multi-pass nature. The weld process model enables identifying the dimensions and characteristics of the three weld regions: base metal, heat-affected zone, and weld metal, and these are treated distinctively. This is followed by virtual fracture experiments, which reveal a very good agreement with laboratory studies. Then, weld pipeline integrity is assessed, estimating critical failure pressures for a wide range of scenarios. Of particular interest is to assess the structural integrity implications of welding defects present in existing natural gas pipelines under consideration for hydrogen transport: pores, lack of penetration, imperfections, lack of fusion, root contraction, and undercutting. The results obtained in hydrogen-containing environments reveal an important role of the weld microstructure and the detrimental effect of weld defects that are likely to be present in existing natural gas pipelines, as they are considered safe in gas pipeline standards.\\ 
\end{abstract}


\end{frontmatter}



\section{Introduction}
\label{Introduction}

In recent years, the global push towards sustainable and renewable energy sources has intensified, with hydrogen emerging as a promising candidate for clean energy storage and transmission \cite{van2020hydrogen}. As a result, efforts have been allocated to establishing a hydrogen energy storage and transport infrastructure \cite{yu2024hydrogen}. However, these efforts have been hindered by the phenomenon of \emph{hydrogen embrittlement}, whereby the ductility, fracture toughness, and fatigue crack growth resistance of metallic materials are very significantly degraded in the presence of hydrogen \cite{Gangloff2003,harris2025hydrogen,chen2024hydrogen}. For example, as little as 6 part per million (ppm) of hydrogen (in weight) can bring down the fracture toughness of pressure vessel steels by an order of magnitude, from above 200 MPa$\sqrt{\text{m}}$ (in the absence of hydrogen) to around 20 MPa$\sqrt{\text{m}}$ \cite{iwadate1977prediction,hosseini2017trapping}.\\

One aspect that has received particular attention is the structural integrity of welds, as these constitute the most susceptible locations in hydrogen storage and transmission infrastructure. During the welding process, microstructural transformations occur due to thermal cycling, resulting in three distinct regions within the welded joint: the weld metal (WM), the heat-affected zone (HAZ), and the base metal (BM), each exhibiting different mechanical properties and fracture behaviors, as well as varying susceptibility to hydrogen embrittlement \cite{ronevich2015hydrogen,diaz2021influence}. The HAZ is of particular interest, as it often contains brittle, hard regions and is therefore the most susceptible element of the weld, setting the limits for component life and performance \cite{Chalfoun2025}. Consequently, experimental studies have been conducted to characterise the behaviour of the HAZ and its susceptibility to hydrogen embrittlement \cite{alvaro2014hydrogen,Ronevich2021}. However, the presence of residual stresses \cite{ronevich2018fatigue,tekkaya2024multi}, the slanted orientation of the HAZ \cite{Bortot2024InvestigationEnvironment,zhou2024multiscale}, and the need for sufficiently large samples to ensure the validity of fracture tests \cite{cui2025fracture}, make the characterisation of weld regions and weld integrity a complex challenge. Moreover, current plans involve retrofitting existing natural gas pipelines to transport hydrogen, bringing in a vast number of scenarios: a wide range of pre-existing defects and numerous combinations of weld and base materials, in addition to any design choice of hydrogen purity and pressure. Computational modelling techniques can be used to overcome experimental constraints and assess the structural integrity of welds in hydrogen environments under a wide range of conditions. \\

Several numerical schemes have been adopted to simulate the coupled deformation-diffusion-fracture behaviour of materials exposed to hydrogen. Examples include continuum damage mechanics \cite{seo2022fracture,kim2024fracture}, cohesive zone modelling \cite{Serebrinsky2004,yu2016uniform,DelBusto2017}, and phase field fracture \cite{Martinez-Paneda2018,diddige2025phase,tan2025phase} approaches. Phase field models for hydrogen embrittlement have gained particular interest due to the link between phase field fracture and well-established fracture mechanics theory \cite{Kristensen2021,yin2025diffusive}, and the numerical robustness of this modelling strategy \cite{kristensen2020phase,mandal2021fracture}, being able to capture complex cracking trajectories and phenomena (crack branching, coalescence) in both 2D and 3D \cite{kristensen2020applications,hai2024novel,yi2024phase}. Phase field models for hydrogen embrittlement have been recently applied to assess the structural integrity of hydrogen transmission pipelines. Zhao and Cheng \cite{zhao2024phase} used a phase field-based model to determine the threshold values of dent depth and internal pressure that will result in hydrogen-assisted failures. Very recently (in 2025), Mandal \textit{et al.} \cite{mandal2024computational} and Wijnen et al. \cite{joboxford,wijnen2025virtual}, combined phase field modelling with thermo-metallurgical welding simulations to predict the critical hydrogen pressures at which pipeline welds would fail. However, they did not explicitly account for trapping effects and limited their study to seam welds. In this work, a new computational framework involving thermo-mechanical welding simulations, elastic-plastic deformation, multi-trap hydrogen diffusion, and hydrogen-sensitive phase field fracture is presented and applied to predict the structural integrity of welds in hydrogen transmission pipelines. Unlike the existing literature, distinct properties are assigned to the various regions of the weld, a wide range of defects (not only crack-like) are considered, and the structural integrity of girth welds is, for the first time, examined. Girth welds are of particular interest as they are conducted on-site and require multiple passes, leading to more complex conditions and microstructural heterogeneity. \\

The remainder of this manuscript is organised as follows. The numerical modelling of the welding process is presented in Section \ref{Sec:WeldingPart}, including details of the numerical framework, the material temperature dependence, girth weld process simulation, thermo-mechanical analysis, and resulting residual stress distributions. Next, in Section \ref{Sec:FracturePart}, the analysis associated with the coupled (deformation-diffusion-fracture) predictions of girth weld integrity are presented. This includes details of the underlying theory, combining elastic-plastic phase field fracture and multi-trap hydrogen diffusion, the numerical implementation, and the results obtained, showcasing the ability of the model to replicate fracture experiments and predict the failure of pipelines containing defects. Of particular emphasis here is the behaviour of pipelines with weld defects that are allowed in the standards (porosity, lack of penetration, imperfections, lack of fusion, root contraction, undercutting) and therefore expected to be present in the existing natural gas pipeline network. Finally, concluding remarks are given in Section \ref{Sec:Conclusions}. 

\section{Welding process modeling}
\label{Sec:WeldingPart}

We begin by presenting the modelling of the welding process, particularising the analysis to girth welds in X80 steel pipelines fabricated through shielded metal arc welding (SMAW). The numerical model is described in Section \ref{Sec:Numerical framework}, while the results obtained are discussed in Section \ref{Sec:ResultsWeld}.

\subsection{Numerical framework}
\label{Sec:Numerical framework}
 
The welding process is modelled as a coupled thermo-mechanical problem. Since the mechanical fields do not influence the thermal problem, this one-way coupled problem is solved in a sequential way, where the temperature field is first computed from the heat transfer equilibrium equation and then transferred as input to the mechanical model. This enables the simulation of the thermal cycles, thermal strains, and residual stresses induced by the multi-pass welding of steel pipelines.
 
 \subsubsection{Governing equations}
\label{Sec:Governing equations}

The spatial and temporal evolution of the temperature \( T \) is governed by the heat equation over the domain \( \Omega \), 
\begin{equation}
\rho(T)\,c(T)\,\frac{\partial T}{\partial t} = \nabla \cdot (k(T) \nabla T) \quad \text{in } \Omega,
\end{equation}

\noindent where \( \rho \) is the mass density, \( c \) is the specific heat capacity, and \( k \) is the thermal conductivity, all of which are temperature-dependent. Heat losses due to convection and radiation are accounted for through the following Neumann boundary condition on \( \partial\Omega \):
\begin{equation}
- k(T) \nabla T \cdot \mathbf{n} = q_c + q_r \quad \text{on } \partial\Omega,
\end{equation}

\noindent where \( \mathbf{n} \) is the unit outward normal to the boundary \( \partial\Omega \). Convective heat transfer is modeled using Newton’s law, 
\begin{equation}
\label{eq:convective}
    q_c = h_c \left( T - T_0 \right) \, ,
\end{equation}
\noindent where $h_c$ is the heat transfer coefficient (equal to $25\,\text{W/m}^2\text{K}$ for steels) and $T_0 = 21^\circ\text{C}$ is the ambient temperature. Radiative heat loss is computed as,
\begin{equation}
\label{eq:radiation}
    q_r = \varepsilon_0 \sigma_0 \left[ (T - T_{\text{abs}})^4 - (T_0 - T_{\text{abs}})^4 \right] \, ,
\end{equation}
\noindent where \( \varepsilon_0 = 0.8 \) is the emissivity, \( \sigma_0 = 5.67 \times 10^{-8} \,\text{W/m}^2\text{K}^4 \) is the Stefan–Boltzmann constant, and \( T_{\text{abs}} = -273^\circ\text{C} \) is the absolute zero temperature.\\ 

The mechanical response of the material is evaluated under the assumption of small strains. In this context, the total strain tensor \( \boldsymbol{\varepsilon} \) is defined as a function of the displacement vector $\mathbf{u}$ as
\begin{equation}
\boldsymbol{\varepsilon} = \frac{1}{2} \left( \nabla \mathbf{u} + \nabla \mathbf{u}^T \right),
\end{equation}
and is additively decomposed into elastic, plastic, and thermal contributions, such that:
\begin{equation}
\boldsymbol{\varepsilon} = \boldsymbol{\varepsilon}^e + \boldsymbol{\varepsilon}^p + \boldsymbol{\varepsilon}^{\text{th}}.
\end{equation}
The thermal strain component is driven by the temperature field $T$, computed in the prior thermal analysis, and is calculated as:
\begin{equation}
\boldsymbol{\varepsilon}^{\text{th}} = \alpha(T) \, (T - T_0) \, \mathbf{I},
\end{equation}
where \( \alpha(T) \) is the temperature-dependent coefficient of thermal expansion and \( \mathbf{I} \) is the identity tensor. Plastic behaviour is estimated using conventional von Mises $J_2$ flow theory, with the evolution of the yield stress being given by the following temperature-dependent isotropic hardening law,
\begin{equation}\label{eq:Powerlaw}
\sigma_y(T,\varepsilon_p) = \sigma_{y0}(T) \left(1 + \frac{E(T) \varepsilon_p}{\sigma_{y0}(T)}\right)^N \, .
\end{equation}
\noindent Here, $\varepsilon_p$ is the equivalent plastic strain, $\sigma_{y0}(T)$ is the initial yield strength at temperature $T$, $E(T)$ is Young’s modulus, and $N$ is the strain hardening exponent. Finally, the stress–strain relationship is defined using the temperature-dependent fourth-order elasticity tensor $\mathbb{E}(T)$, 
\begin{equation}
\boldsymbol{\sigma} = \mathbb{E}(T) : (\boldsymbol{\varepsilon} - \boldsymbol{\varepsilon}^p - \boldsymbol{\varepsilon}^{\text{th}}) \, .
\end{equation}

\subsubsection{Material property variation with temperature}
\label{Sec:Material temperature dependence}

\begin{figure}[H]
    \centering
    \includegraphics[width=0.99\textwidth]{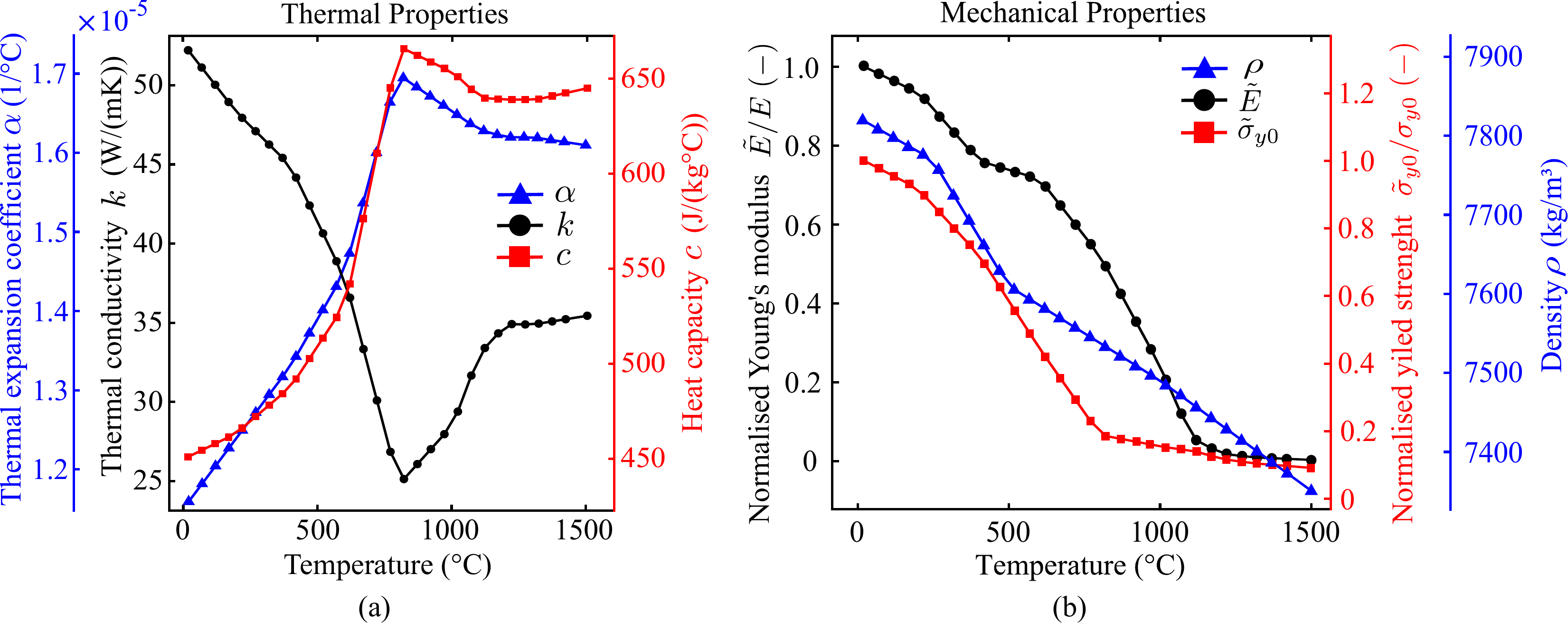}
    \caption{Temperature dependence of relevant material properties. (a) Thermal properties: coefficient of thermal expansion (\(\alpha\)), thermal conductivity (\(k\)), and specific heat (\(c\)). (b) Mechanical properties: Young's modulus ($E$), initial yield strength ($\sigma_{y0}$) and density $\rho$; a tilde is used to denote the temperature-dependent parameter.}
    \label{fig:MaterialProps}
\end{figure}

We proceed to define how the relevant thermal and material properties depend on temperature. To this end, the general framework described in Section \ref{Sec:Governing equations} is particularised to the analysis of X80 pipeline steel, a material extensively used in the natural gas pipeline network that is being considered for hydrogen transport. The thermal and mechanical properties of X80 steel were obtained from experimental data in the literature \cite{Xu2024,Huang2024}. The temperature dependency of the relevant thermal properties (thermal expansion coefficient $\alpha$, heat capacity $c$, thermal conductivity $k$) is given in Fig. \ref{fig:MaterialProps}a, over a range going from room temperature to 1500$^\circ$ C, the assumed melting point of the material in the simulations. The variation with temperature of the thermal properties given in Fig. \ref{fig:MaterialProps}a follows the typical behaviour of pipeline steels. The temperature dependency of relevant mechanical (Young's modulus $E$, initial yield stress $\sigma_{y0}$) and physical (density, $\rho$) properties is given in Fig. \ref{fig:MaterialProps}b. Again, the changes with temperature presented follow the expected pattern for pipeline steels, with the density showing a small sensitivity, while Young's modulus and yield strength experience a more significant temperature dependence. The same sensitivity to changes in temperature is assumed for the weld (filler) material, with the only difference being the room temperature value of Young's modulus (180300 MPa) and initial yield strength (688 MPa), to ensure consistency with experimental measurements \cite{Yang2015}. The Poisson’s ratio was assumed to remain constant with temperature and set to 0.3 for both base and weld material. Similarly, the strain hardening coefficient $N$ was considered constant with temperature, with values of 0.1 for the BM and 0.07 for the WM, according to the uniaxial stress–strain data reported in Ref. \cite{Yang2015}.\\

\subsubsection{Welding model}
\label{Sec:Thermo-Mechanical analysis}

\begin{figure}[H]
    \centering
    \includegraphics[width=0.85
    \textwidth]{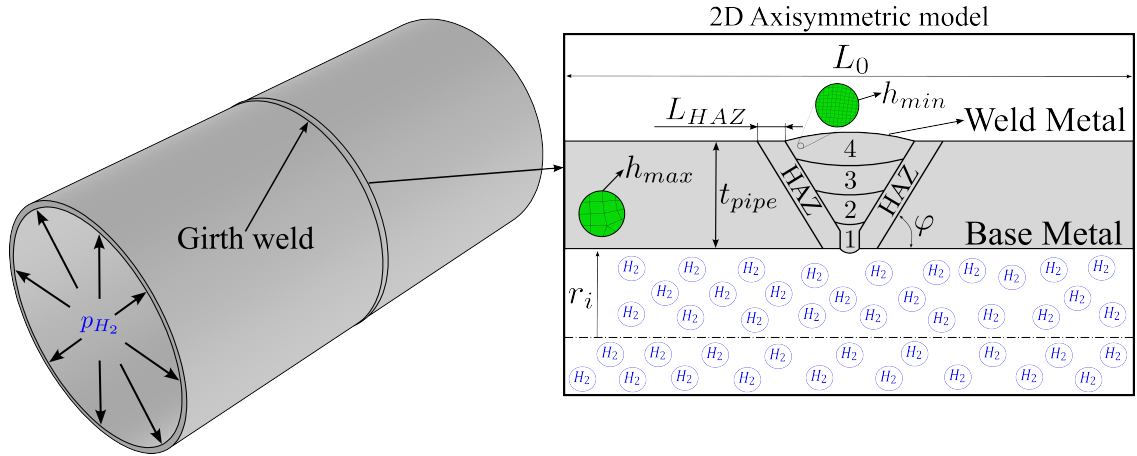}
    \caption{Schematic representation of the boundary value problem considered: a 4-pass girth weld, with weld angle $\varphi$ and HAZ width $L_{\text{HAZ}}$, in a pipeline with inner radius $r_i$ and thickness $t_{\text{pipe}}$. The pipeline is made of X80 pipeline steel and assumed to transport hydrogen at a pressure $p_{H_2}$. A 2D axisymmetric model is employed, with a gradually refined mesh with a maximum element size of $h_{\text{max}}$, in the edges of the domain, and a minimum element size of $h_{\text{min}}$, in the weld region. The longitudinal axis of the pipeline, which corresponds to the axis of symmetry in the model, is represented with a dashed line.}
    \label{fig:WeldGeom}
\end{figure}

The weld process model is particularised to the analysis of a multi-pass girth weld. As shown in Fig. \ref{fig:WeldGeom}, we model a 4-bead weld on a pipeline with an inner radius of $r_i = 228$ mm and a thickness of $t_{pipe} = 12$ mm. These dimensions meet the API pipeline design specifications \cite{AmericanPetroleumInstitute2018}. The weld angle $\varphi$ is taken to be 60$^\circ$, a common value for this type of weld \cite{Mathias2013}. Taking advantage of the inherent axial symmetry of girth welds, a 2D axisymmetric model is employed. The width of the modelled domain is taken to be \( L_0 = 120 \) mm, a sufficiently large value to prevent edge effects. Two materials are considered in the welding simulation process: the BM and the WM, with their thermo-mechanical properties being given in Section \ref{Sec:Material temperature dependence}. The weld process simulations will allow establishing a third domain, the HAZ, for the structural integrity analysis presented in Section \ref{Sec:FracturePart}, where its properties and behaviour will be discussed. In this vein, the thermo-mechanical weld process results will dictate the width of the heat-affected zone ($L_{\text{HAZ}}$), its characteristic dimension, as discussed below (Section \ref{Sec:ResultsWeld}).

The domain is discretised using quadratic finite elements, with the mesh gradually refining towards the centre of the domain (WM region). Following a mesh sensitivity study, the largest elements are located in the edges of the model (BM), where the characteristic element size is defined to be $h_{max} = 1.5$ mm, while the minimum element size is located in the welded area, with a characteristic size of $ h_{min}= 0.02$ mm. A total of 186,110 elements are employed. The model is implemented in the commercial finite element package \texttt{Abaqus}. A \texttt{UMAT} subroutine is used to define the thermo-elastic-plastic behaviour described in Section \ref{Sec:Governing equations}. To facilitate the simulation of the welding process, the add-on \texttt{AWI} (Abaqus Welding Interface) is used. Typical calculation times are of 30 to 40 minutes in a standard workstation (Intel(R) Xeon(R) Gold 6242R CPU @ 3.10 GHz). The electrode welding process is modelled as a sequence of steps, typically denoted as:
\begin{itemize}
    \item \emph{Apply Torch} step. During this stage, the temperature at the weld cavity edges is progressively increased to the specified melting point of 1500$^\circ$C over eight seconds.
    \item \emph{Hold Torch} step. This step determines the heat input and governs the temporal 
    evolution of the temperature field in the model. The temperature at the cavity edges is 
    maintained at 1500$^\circ$C to allow heat penetration into the material. In this work, a 
    holding time of 4~s was selected, calibrated in the first welding pass by comparing the 
    simulated thermal cycle with experimental measurements obtained from a single-pass of X80 
    girth weld \cite{Gaudet2015}.
    \item \emph{Pause Torch} step. At this point, the elements within the weld bead, initially at 1500$^\circ$C, are activated in the model, allowing heat to diffuse through the domain. This is a very fast, instantaneous step, serving only as a transitional stage; a step duration of \(1.0 \times 10^{-7}\,\text{s}\) is considered here.
    \item \emph{Cool-down} step. This step simulates the inter-pass cooling periods, during which the heat from each weld pass dissipates through the component until a target inter-pass temperature of 125~$^\circ$C is reached, replicating the values adopted in Ref. \cite{Yu2023}. The duration of each cool-down is controlled by the \texttt{UAMP} subroutine, which terminates the step once the temperature at the weld bead reaches the specified target. For all weld passes except the final one, cooling was applied until a temperature of 125~$^\circ$C was reached. For the last pass, the cooling continued down to ambient temperature (21~$^\circ$C), completing the thermal cycle.
\end{itemize}

This protocol enables capturing the evolution of thermal and mechanical fields within a cross-section of the weld, simplifying a 3D problem \cite{Seles,Parmar2016}. The boundary conditions associated with the thermal and mechanical models are defined as follows. In the thermal model, the convective and radiative heat fluxes, \(q_c\) and \(q_r\), are applied on all relevant external surfaces of the model, including the outer and inner surfaces of the pipe, as well as at the surfaces of the deposited weld beads exposed to air, as defined in Eqs.~(\ref{eq:convective}) and~(\ref{eq:radiation}), respectively. The Dirichlet thermal boundary conditions are imposed during the \textit{Apply Torch} and \textit{Hold Torch} steps by prescribing a temperature on the cavity corresponding to each weld bead; the sensor nodes that control the duration of the \textit{Hold Torch} and \textit{Cool-down} steps are automatically generated by the \texttt{AWI} add-on. In the mechanical model, the longitudinal displacements (\(u_l\)) of the lateral edges are restricted to replicate the clamping conditions applied during the welding process, which prevent longitudinal expansion of the BM. Additionally, the condition $u_r = 0$ is imposed along the axis of revolution to satisfy the rigid body condition. These boundary conditions are applied throughout all simulation steps, starting with the thermal analysis to compute the temperature distribution over time. Then, the resulting temperature field is introduced into the mechanical analysis by defining a \texttt{Predefined Temperature Field} in \texttt{Abaqus}, which assigns the temperature history to the mechanical model. In this way, a sequentially coupled thermo-mechanical analysis is performed, where the thermal field drives the temperature-dependent mechanical response described in Section~\ref{Sec:Governing equations}, which is defined through the user material (\texttt{UMAT}) subroutine.

\subsection{Results}
\label{Sec:ResultsWeld}

\begin{figure}[H]
    \centering
    \includegraphics[width=1\textwidth]{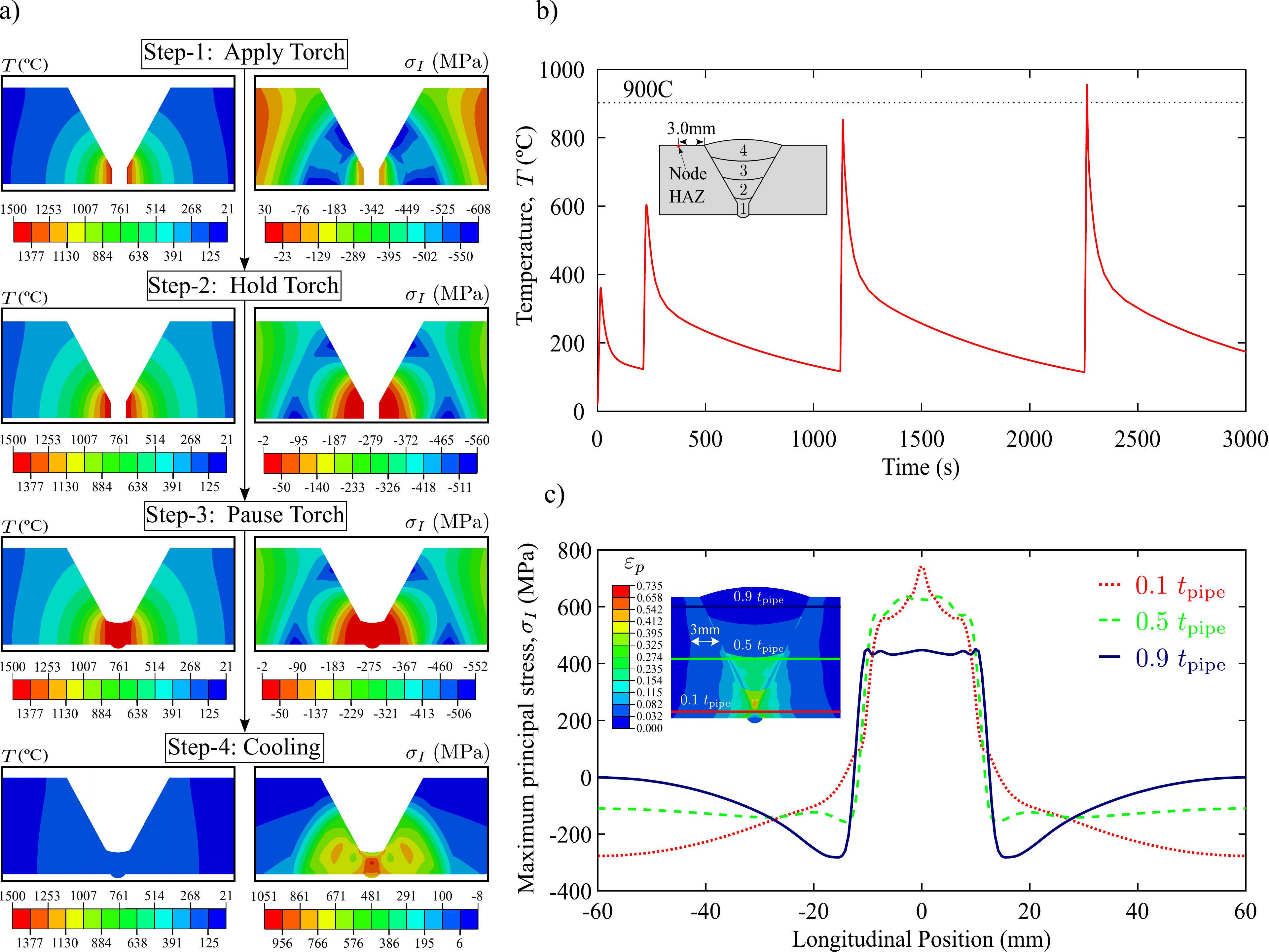}
    \caption{Welding process modelling and resulting fields. 
(a) Sequence of steps for the \textit{first welding pass}, showing the evolution of thermal profiles and maximum principal stresses ($\sigma_I$): 
(Step~1) Apply Torch, (Step~2) Hold Torch, (Step~3) Pause Torch, and (Step~4) Cool-down. 
(b) Evolution of the temperature in a node located in the heat-affected zone (HAZ) at a distance of 3~mm from the fusion line. 
(c) Residual stress distribution at the end of the welding process, represented by contours of equivalent plastic strain ($\varepsilon_{p}$) and maximum principal stress ($\sigma_I$) values along the weld longitudinal direction, sampled at different relative depths of the pipe thickness: 0.1$t_{\text{pipe}}$, 0.5$t_{\text{pipe}}$, and 0.9$t_{\text{pipe}}$.}
    \label{fig:WeldingResults}
\end{figure}

We proceed to discuss the insight gained in the thermo-mechanical welding simulations, which is also the starting point for the subsequent deformation-diffusion-fracture modelling analysis (Section \ref{Sec:FracturePart}). First, Fig.~\ref{fig:WeldingResults}a illustrates the evolution of both the temperature profiles and the maximum principal stresses ($\sigma_I$, corresponding to the circumferential stresses $\sigma_\theta$) experienced in the welded region during the first pass, representative of the thermo-mechanical response observed in the subsequent passes. Initially, the electrode is applied (\emph{Apply Torch} step), progressively raising the temperature within the bead cavity to the melting point of the metal ($T = 1500 \, ^\circ \mathrm{C}$), while compressive stresses develop in the surrounding regions due to the thermal expansion constraint. Then, the electrode is maintained at the target temperature of $T = 1500 \, ^\circ \mathrm{C}$ (\emph{Hold Torch} step). This results in vanishing stresses in the weld bead, consistent with the loss of stiffness associated with temperatures closer to the melting point (see Fig.~\ref{fig:MaterialProps}b). Following an instantaneous \emph{Pause Torch} step, the bead is activated with an initial temperature of 1500~$^\circ$C, and the temperature decreases as part of the \emph{Cool-down} step until the interpass temperature of 125~$^\circ$C is reached, during which the largest residual stresses are generated because the thermal contraction of the WM is constrained by the surrounding BM, preventing free shrinkage and leading to the development of tensile stresses in the weld region. At this point, the next weld pass begins, following the same sequence of steps. Overall, the thermo-mechanical analysis highlights the typical temperature distribution and stress development in multi-pass welding, which are further detailed through the thermal cycle of a representative HAZ node (Fig. \ref{fig:WeldingResults}b) and the residual stress fields at the end of the process (Fig. \ref{fig:WeldingResults}c). 

The temperature evolution of a representative HAZ node located at the outer surface of the pipe during the four weld passes is shown in Fig.~\ref{fig:WeldingResults}b. The results show how the peak temperature increases with each pass, due to the node's location. The cooling stage after each of the first three passes proceeds down to the interpass temperature of 125~$^\circ$C, while after the fourth pass, the cooling continues until room temperature is reached.
In our simulations, the width of the HAZ is defined according to the method proposed by Garcin et al. \cite{Garcin2016}, based on the distance from the fusion line to the points where the peak temperature exceeds 900$^\circ$C. Following this definition, the HAZ in our model spans approximately 3 mm, as shown in Fig.~\ref{fig:WeldingResults}b . This is well aligned with existing experimental studies \cite{Yang2015,Chalfoun2025}.\\

The resulting residual stresses are assessed in Fig.~\ref{fig:WeldingResults}c, where contours of the equivalent plastic strain $\varepsilon_{p}$ are provided, together with the values of the maximum principal strain $\sigma_I$ along the weld longitudinal direction, sampled at various depths within the pipeline thickness (0.1$t_{\text{pipe}}$, 0.5$t_{\text{pipe}}$, and 0.9$t_{\text{pipe}}$). Consider first the equivalent plastic strain contours, which reveal a greater accumulation of plastic strains at the weld root due to the intense thermal gradients and local constraint generated during the initial passes. This localised plastic deformation leads to a higher residual stress concentration, as confirmed by the distribution of $\sigma_I$ along the 0.1$t_{\text{pipe}}$ path. In this region, peak tensile residual stresses reach values of about 730~MPa, thus exceeding the yield strength of the WM (688 MPa), as a result of the strong restraint imposed by the surrounding BM. Near the weld cap (0.9$t_{\text{pipe}}$), lower maximum principal stresses are observed. This reduction is attributed to the fact that the upper layers are deposited later, once the previously deposited WM has already undergone thermal and mechanical relaxation, allowing for stress redistribution. Additionally, the upper part of the weld is less constrained and experiences less straining, resulting in lower residual stress levels (around 450 MPa) that still remain tensile.
At mid-thickness (0.5$t_{\text{pipe}}$), residual stresses exhibit intermediate peak values in the welded zone (about 600 MPa), reflecting a stress gradient developed through the thickness as a consequence of the sequential thermal loading and plastic strain accumulation associated with the multi-pass welding process. In addition, the mechanical results also highlight the presence of a roughly 3 mm wide region adjacent to the fusion line, where plastic deformation accumulates and higher residual stresses are attained (relative to the BM), strengthening our definition of the HAZ region.

\section{Coupled deformation-diffusion-fracture predictions of girth weld integrity}
\label{Sec:FracturePart}

The thermo-mechanical weld process simulation provides the starting point for the structural integrity predictions, defining the residual stress distribution and the various weld regions. In this Section, the coupled structural integrity model is presented, and its ability to reproduce experiments and deliver service life predictions is assessed. We begin by describing the numerical model in Section \ref{Sec:CoupledModel}, including details of the underlying theory and numerical implementation. Then, in Section \ref{sec:Validation}, we benchmark the model against fracture experiments on X80 BM, WM, and HAZ samples. The validated model is then employed in Section \ref{Sec:ResultsFracture} to study the structural integrity of hydrogen transmission pipelines containing a wide range of defects. 

\subsection{Modelling framework}
\label{Sec:CoupledModel}

A coupled theory for hydrogen transport (including trapping), elastic-plastic mechanical deformation, and hydrogen-assisted phase field fracture is presented. The work builds upon Refs. \cite{isfandbod2021mechanism,cupertino2024suitability}, but brings additional elements of novelty, such as the coupling between a thermodynamically-consistent driving force and the consideration of multiple trapping sites, including trap creation. The primary variables of the coupled formulation are the displacement vector $\mathbf{u}$, the lattice hydrogen concentration $C_L$, and the phase field order parameter $\phi$. The coupled balance equations for, respectively, the deformation, hydrogen diffusion, and fracture sub-problems are given by,
\begin{equation}\label{eq:MechDef}
\nabla \cdot \boldsymbol{\sigma} = 0, \quad 
\text{where} \quad \boldsymbol{\sigma} = g(\phi) \mathbb{E} : (\boldsymbol{\varepsilon} - \boldsymbol{\varepsilon}^p), 
\quad \text{and} \quad |\boldsymbol{\sigma}_{\text{dev}}| = g_p(\phi) \sigma_y(\varepsilon_{p}), 
\quad \text{if} \quad \dot{\varepsilon}_{p} > 0, 
\end{equation}
\begin{align}\label{eq:Htransport}
\frac{D_L}{D_e} \frac{\partial C_L}{\partial t} + \theta_T \frac{\text{d} N_T^{(d)}}{\text{d} \varepsilon_p} \frac{\text{d} \varepsilon_p}{\text{d} t} = {D_L} \nabla^2 C_L & - \nabla \cdot \left( \frac{{D_L} C_L}{RT} V_H \nabla \sigma_H \right), \nonumber \\
& \text{where} \,\,\,\, D_e =   \frac{D_L C_L}{C_L + \sum_{i} C_T^{(i)} (1 - \theta_T^{(i)})}, 
\end{align}
\begin{equation}\label{eq:PF}
G_c (C_L) \left( \frac{\phi}{\ell} - \ell \nabla^2 \phi \right) = 2(1 - \phi) \mathcal{H},  \quad 
\text{where} 
\quad \mathcal{H} := \max_{\tau \in [0,t]} \left\{ \psi_e^{+}(\tau) \right\} + 0.1 \psi_p \, .
\end{equation} \vspace{3pt}

Equation (\ref{eq:MechDef}) is the balance of linear momentum for an elastic-plastic solid deformed under quasi-static conditions and undergoing damage. Small strains are assumed, with the total strain tensor additively decomposing into its elastic and plastic parts: $\bm{\varepsilon}=\bm{\varepsilon}^e+\bm{\varepsilon}^p$. The role of cracking in reducing the load carrying capacity is captured by the use of a degradation function $g (\phi)$, defined below, which multiplies the fourth-order elasticity tensor $\mathbb{E}$. Elastic-plastic behaviour for a solid with yield stress $\sigma_y$ is defined based on von Mises ($J_2$) plasticity theory, defining a consistent coupling between damage and plasticity, as described in Section \ref{Sec:PFmodel}.\\

Equation (\ref{eq:Htransport}) describes the hydrogen transport process, through an effective diffusivity ($D_e$), two-layer approach \cite{krom1999hydrogen,diaz2024explicit}, whereby the hydrogen concentration can be additively decomposed into two parts: the hydrogen concentration at lattice sites, $C_L$, and the hydrogen concentration at trapping sites, $C_T$. The lattice and trap occupancies can be subsequently defined as $\theta_L= C_L/N_L$ and $\theta_T=C_T/N_T$, with $N_L$ and $N_T$ being the density of lattice and trapping sites, respectively. Multiple trap types can be accommodated, as described with the $(i)$ superscript (corresponding to trap type $i$), with the constitutive details of the multi-trapping treatment being discussed below. As defined in detail in Section \ref{Sec:HtransportModel}, dislocation traps can evolve, and this is modelled through the `Krom term' \cite{krom1999hydrogen}, involving the dislocation trap density $N_T^{(d)}$ and the equivalent plastic strain $\varepsilon_p$. Eq. (\ref{eq:Htransport}) captures the role of traps in slowing hydrogen diffusion \cite{cupertino2023hydrogen}, with $D_e$ being lower than the lattice diffusion coefficient $D_L$, as well as the role of lattice expansion in leading to higher hydrogen contents in highly stressed areas, through the term involving the partial molar volume of hydrogen $V_H$ and the hydrostatic stress $\sigma_H$.\\

Equation (\ref{eq:PF}) describes the fracture process, through the evolution of the phase field order parameter $\phi$. Consistent with thermodynamics, cracks are assumed to grow if the energy stored in the solid, characterised by the history field $\mathcal{H}$, reaches a critical quantity - the material toughness $G_c$. The latter is defined to be dependent on the lattice hydrogen concentration to capture the embrittlement effects of hydrogen, with further details given below. The fracture driving force, $\mathcal{H}$, includes the tensile component of the elastic strain energy density $\psi_e^{+}$ and 10\% of its plastic counterpart, $\psi_p$, as experiments show that 90\% of the plastic work is dissipated into heat and thus not available to generate new cracks \cite{cupertino2024suitability,TaylorQuinney}. As described below (Section \ref{Sec:PFmodel}), the role of the material strength is also accounted for, through the phase field length scale $\ell$.\\

The particularities and constitutive choices for each sub-problem are presented in the sections below, along with their coupling elements.

\subsubsection{Multi-trap hydrogen transport model}
\label{Sec:HtransportModel}

Hydrogen transport is predicted based on a two-level (lattice-trap) effective diffusivity model, based on Oriani's thermodynamic equilibrium \cite{R.A.Oriani1974}, as presented in Eq. (\ref{eq:Htransport}). Accordingly, trapping kinetics are assumed to be much faster than diffusion kinetics (a sensible assumption in most scenarios \cite{garcia2024tds}), and this leads to the following Fermi-Dirac relationship between the occupancy of lattice sites ($\theta_L$) and the occupancy of trapping sites of type $i$ ($\theta_T^{(i)}$),
\begin{equation}\label{Eq:Oriani}
\frac{\theta_T^{(i)}}{1-\theta_T^{(i)}} = \frac{\theta_L}{1-\theta_L} \cdot \exp{\left(\frac{-W_B^{(i)}}{RT}\right)} \, ,
\end{equation}
\noindent where $W_B^{(i)}$ is the binding energy of each trap, and $R$ is the gas constant. Hence, Eq. (\ref{Eq:Oriani}) relates, at the integration point level, each $C_L$ value (and associated $\theta_L$) to the trap occupancy $\theta_T$ and concentration $C_T$, which in turn define the effective diffusivity - see Eq. (\ref{eq:Htransport}).\\

Our framework is particularised to the study of pipeline steels, with the case studies addressing the failure of welded X80 pipeline components. Accordingly, four primary types of hydrogen traps are considered: dislocations, grain boundaries, cementite-ferrite interfaces ($\alpha-\mathrm{Fe}_3\mathrm{C}$), and martensite-austenite (M/A) interfaces \cite{Zheng2023,Sun2021}. The trap densities of grain boundaries, $\alpha-\mathrm{Fe}_3\mathrm{C}$, and martensite-austenite interfaces remain constant throughout the analysis, but this is not the case for dislocations, which evolve with plastic deformation. Several phenomenological expressions have been proposed to relate the equivalent plastic strain ($\varepsilon_p$) to the dislocation density ($\rho$) \cite{diaz2016coupled,jemblie2017review,fernandez2020analysis}; here, we follow Ref. \cite{LUFRANO1998827} and define $N^{d}_{T}$ to follow a geometrical relationship with the dislocation density ($\rho$), which for a bcc lattice is given by,
\begin{equation}\label{eq:NT}
N_T^{d} = \frac{\sqrt{2} \rho}{d} \, ,
\end{equation}
where $d$ is the lattice parameter ($2.866 \times 10^{-10}$ m for iron). For simplicity, only statistically stored dislocations (SSDs) are considered, although geometrically necessary dislocations (GNDs) can dominate ahead of stress concentrators \cite{martinez2016strain}. Accordingly, the dislocation density is assumed to evolve through a piece-wise dependency with the equivalent plastic strain \cite{Gilman1969}:
\begin{equation}\label{eq:rhossd}
\rho =
\begin{cases}
\rho_0 + 2 \gamma \varepsilon_p & \text{if } \varepsilon_{p}\leq 0.5 \\
\rho_0 + \gamma & \text{if } \varepsilon_{p} > 0.5
\end{cases}
\end{equation}

\noindent where \(\rho_0 = 10^{10} \, \text{m}^{-2}\) is the dislocation density in the unstrained condition and \(\gamma = 10^{16} \, \text{m}^{-2}\) represents an experimentally calibrated quantity, as reported in Ref.~\cite{dadf2011}. Considering Eq. (\ref{eq:NT}) and the dislocation density evolution, Eq. (\ref{eq:rhossd}), the rate of trap creation term in Eq. (\ref{eq:Htransport}) is given by
\begin{equation}
\frac{dN_T^{dis}}{d\varepsilon_p} =
\begin{cases}
\sqrt{2} \gamma / d & \text{if } \varepsilon_p < 0.5 \\
0 & \text{if } \varepsilon_p \geq 0.5
\end{cases}
\end{equation}

As shown below (Section \ref{Sec:ResultsFracture}), the numerical experiments capture how the trap density evolves, and how this impacts hydrogen diffusion and trapping. Quantitative choices for the trap density and binding energies for each trap type are provided in Table \ref{tab:binding_energies}, based on experimental measurements from the literature \cite{Shang2024,Li2020,Zheng2023,H.K.D.H.Bhadeshia2016}. The structural integrity analyses consider three distinct weld regions:BM,WM, and HAZ, each of which has different material properties. Accordingly, different trap densities are considered in each of these regions, capturing the heterogeneous nature of hydrogen trapping across the weld. These are defined through a combination of literature data, microstructural considerations, and 1D numerical permeation experiments. Martensite/austenite (M/A) constituents are scarce in the BM but form more extensively in the WM and HAZ regions due to thermal cycling and microstructural transformations \cite{ZHANG2024111764}. Trap sites associated with cementite-ferrite interfaces ($\alpha$-Fe$_3$C), typically found in pearlitic regions, are mainly present in the base metal. The density of grain boundary traps is lowest in the HAZ, as it is the region with coaser grains, with the BM having the smallest grains and thus higher grain boundary trap density. To ensure that the values chosen result in hydrogen transport behaviour in agreement with experiments, hydrogen permeation in membranes is simulated to ensure that the resulting apparent diffusivity agrees with the effective diffusion coefficients that have been measured for each weld region; namely, $D_e^{BM}=2.8 \times 10^{-5}$ mm$^2$/s, $D_e^{HAZ}=2.0 \times 10^{-5}$ mm$^2$/s, and $D_e^{WM}= 1.7\times 10^{-4}$ mm$^2$/s \cite{Yan2014}. In these numerical experiments, the lattice diffusion coefficient $D_L$ has been assumed to be the same across the weld, as the lattice behaviour of the three weld regions can be well described by the diffusivity of the bcc lattice: $D_L = 7.2 \times 10^{-3}\ \mathrm{mm^2/s}$ \cite{garcia2024tds}. Throughout this paper, the lattice density is given by $N_L = 5.2 \times 10^{20}$ sites/mm$^3$, based on the atomic packing of bcc iron at room temperature \cite{fernandez2020analysis}. 

\begin{table}[H]
    \centering
    \begin{tabular}{|l|c|c|c|c|c|}
        \hline
        \textbf{Trap Type} & \( W_B \) (kJ/mol) & \( N_T^{BM} (\mathrm{sites/m^3}
) \) & \( N_T^{HAZ} (\mathrm{sites/m^3}
) \) & \( N_T^{WM} (\mathrm{sites/m^3}
) \) & Ref. \\ \hline
        Dislocations & 25.0 & \(3.13 \times 10^{36}\) & \(4.25 \times 10^{36}\) & \(5.15 \times 10^{35}\) & \cite{Shang2024} \\ \hline
        M/A Interfaces & 47.1 & \(2.56 \times 10^{21}\) & \(9.87 \times 10^{23}\) & \(9.56 \times 10^{21}\) & \cite{Zheng2023} \\ \hline
        $\alpha$-Fe\textsubscript{3}C  & 13.5 & \(9.26 \times 10^{22}\) & \(3.23 \times 10^{22}\) & \(2.58 \times 10^{22}\) & \cite{Zheng2023}  \\ \hline
        Grain Boundaries & 32.0  & \(9.12 \times 10^{21}\) & \(1.63 \times 10^{20}\) & \(5.50 \times 10^{20}\) & \cite{H.K.D.H.Bhadeshia2016} \\ \hline
    \end{tabular}
    \caption{Binding energies \( W_B \) and trap densities \( N_T \) adopted. The weld trapping heterogeneity is defined by considering different trap densities in the BM (\( N_T^{BM} \)), WM (\( N_T^{WM} \)), and HAZ (\( N_T^{HAZ} \)). In all cases, the trap density for dislocations corresponds to the unstrained condition, \( N^{(d)}_{T,0} \).}
    \label{tab:binding_energies}
\end{table}

\subsubsection{Hydrogen-dependent fracture of elastic-plastic solids}
\label{Sec:PFmodel}

Material deformation and fracture are simulated through a hydrogen-sensitive elastic-plastic phase field formulation. Cracks are tracked as diffuse interfaces, using a phase field variable $\phi$ to regularise discrete cracks over a finite domain, with $\phi=1$ representing fully damaged material while $\phi=0$ denotes intact (undamaged) material \cite{bourdin2000numerical}. A smooth transition between these states is achieved through an energy density functional, ensuring a variationally consistent fracture formulation and predicting the evolution of cracks as an exchange between stored and fracture energy. The total free energy functional, defined over the domain of the elasto-plastic solid $\Omega$, is given by:
\begin{equation}\label{freenergy}
\Psi = \int_\Omega \left( \psi + \gamma(\phi) G_c(C_{L}) \right) \,\, \text{d}\Omega \, ,
\end{equation}
where $\psi$ is the strain energy density, $G_c(C_{L})$ is the hydrogen-dependent material toughness, and $\gamma(\phi)$ is the crack regularization function, taken here to be
\begin{equation}
\gamma(\phi) = \frac{1}{2\ell} (\phi^2 + \ell^2 |\nabla \phi|^2),
\end{equation}
with $\ell$ being the length scale parameter that governs the width of the fracture process zone. In the context of the so-called \texttt{AT2} phase field fracture model, adopted here, the phase field length scale can be directly related to the material strength $\sigma_c$ upon some assumptions (see \cite{Martinez-Paneda2018,tanne2018crack,Kristensen2021} and Refs. therein), such that
\begin{equation}\label{eq:lc}
\ell = \frac{27 E G_c(C_{L})}{256 \sigma_c^2} \, ,
\end{equation}
The link to the material strength endows phase field fracture models with the ability to predict crack nucleation and well-established phenomena such as the crack size dependency \cite{tanne2018crack,Kristensen2021,molnar2020toughness}.\\

Following Refs. \cite{mandal2024computational,cupertino2024suitability}, the strain energy density of the elastoplastic solid $\psi$ is defined as:
\begin{equation}
    \psi = g(\phi)\psi^+_e(\boldsymbol{\varepsilon}^e) + \psi^-_e(\boldsymbol{\varepsilon}^e) + g_p(\phi)\psi_p(\varepsilon_{p}) \, ,
\end{equation}
where 
\begin{equation}
    \psi^+_e = \frac{K}{2}\langle \text{tr}\boldsymbol{\varepsilon}^e \rangle^2_+ + G \boldsymbol{\varepsilon}^e_\text{dev} : \boldsymbol{\varepsilon}_\text{dev}^e \, , \quad \psi^-_e = \frac{K}{2}\langle \text{tr}\boldsymbol{\varepsilon}^e \rangle^2_{-} \, , \quad \psi^p = \frac{\sigma_{y0}^2}{E(N+1)} \left( \left(1 + \frac{E \varepsilon_{p}}{\sigma_{y0}} \right)^{(N+1)} - 1 \right) \, .
\end{equation}
Here, $K$ is the bulk modulus, $G$ is the shear modulus, $\langle a \rangle_{\pm}=(a\pm |a|)/2$ and $\bm{\varepsilon}_{\text{dev}}^e=\bm{\varepsilon}^e-\text{tr}(\bm{\varepsilon}^e) \bm{I} /3$.. This formulation incorporates a volumetric-deviatoric split \cite{Amor2009} to prevent crack propagation under compressive stress conditions. Two degradation functions are employed in this study: the elastic degradation function,
\begin{equation}
    g(\phi) = (1 - \phi)^2 \, ,
\end{equation}
\noindent to reduce the material stiffness due to fracture, and the plastic degradation function 
\begin{equation}
    g_p(\phi) = \beta g(\phi) + (1 - \beta), \quad \text{with} \quad 0 \leq \beta \leq 1 \, ,
\end{equation}
\noindent which degrades both the yield surface and the plastic energy. Here, $\beta$ quantifies the fraction of plastic work stored in the material and available for crack propagation; as stated above, a value of $\beta = 0.1$ is adopted, following the seminal experiments by Taylor and Quinney \cite{TaylorQuinney,mandal2024computational}. To ensure damage irreversibility, a fracture driving force $\mathcal{H}$ is defined, as shown in Eq. (\ref{eq:PF}). Considering the aforementioned constitutive assumptions and deriving the governing equations by taking variations of the total free energy functional, Eq. (\ref{freenergy}), with respect to the phase field variable $\phi$ and the displacement field $\mathbf{u}$, the coupled problem of elastic-plastic deformation and fracture is obtained - see Eqs. (\ref{eq:MechDef}) and (\ref{eq:PF}).\\

\begin{figure}[H]
    \centering
    \includegraphics[width=0.9\textwidth]{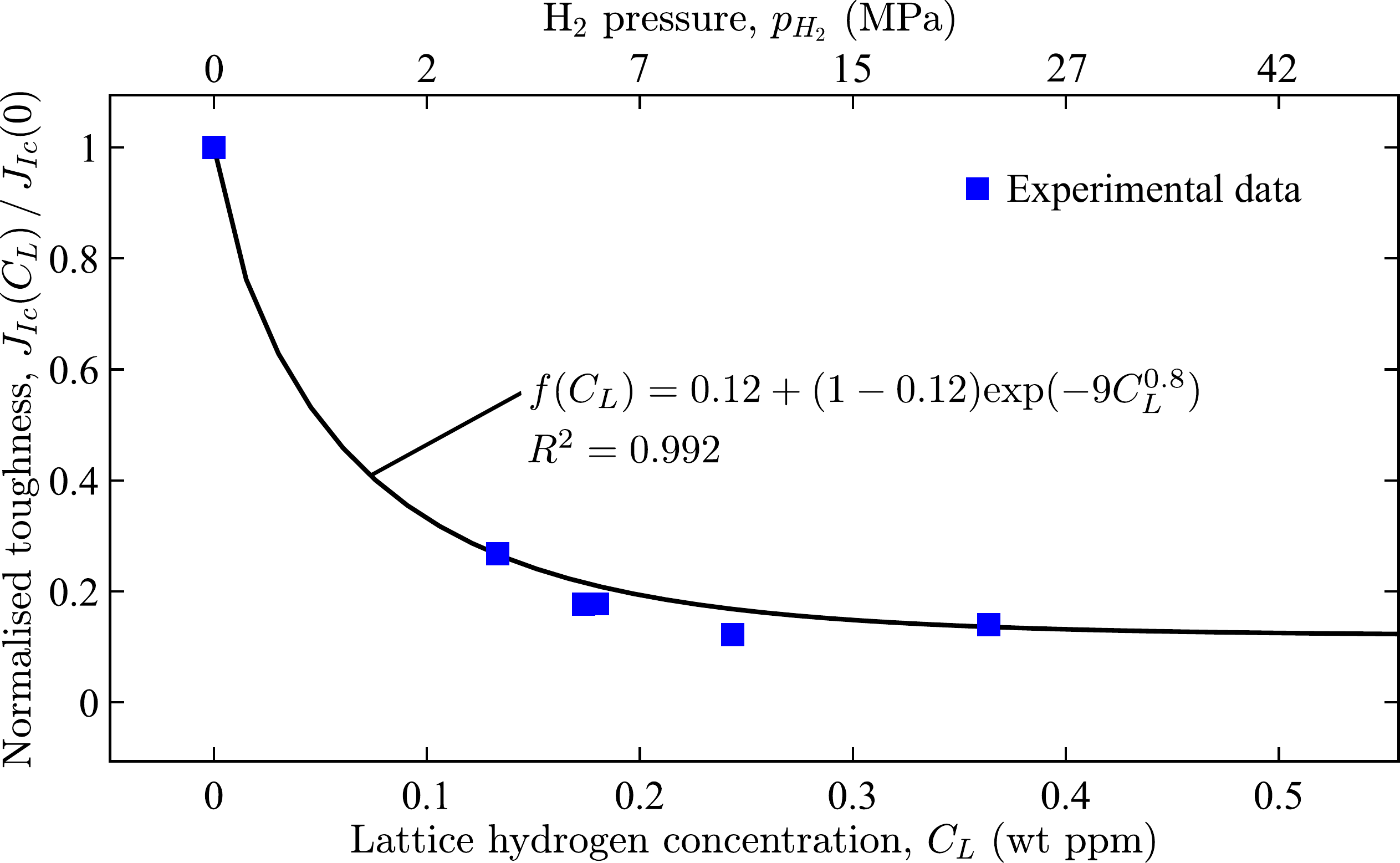}
    \caption{Determining the hydrogen degradation law from experimental toughness versus H$_2$ pressure data from the literature \cite{Marchi2011, Marchi2022, Shang2021}. The normalised toughness versus lattice hydrogen content data is very well approximated ($R^{2}=0.992$) with the degradation law, Eq. (\ref{eq:DegradationLaw}), upon the assumption of the following degradation coefficients: $\xi=0.12$, $\eta=9$, and $b=0.8$.}
    \label{fig:Jic_h}
\end{figure}

One must also define how the fracture energy depends on the hydrogen content. To this end, we follow Mandal \textit{et al.} \cite{mandal2024computational} and assume the following degradation of the fracture energy with the lattice hydrogen content \( C_L \),
\begin{equation} \label{eq:DegradationLaw}
f(C_L) = \frac{G_c(C_L)}{G_c(0)} \equiv \frac{J_{Ic}(C_L)}{J_{Ic}(0)} = 
\xi + (1 - \xi) \exp\left(-\eta C_L^b\right),
\end{equation}
where $\xi$, $\eta$, and $b$ are material-dependent coefficients. This function appropriately captures the decay in toughness with hydrogen content observed in pipeline steels, through appropriate choices of the parameters $\xi$, $\eta$, and $b$. To this end, Fig. \ref{fig:Jic_h} presents a compilation of experimental literature data on toughness versus H$_2$ pressure for X80 pipeline steel \cite{Marchi2011, Marchi2022, Shang2021}. The associated lattice hydrogen concentration is also shown, as obtained from the pressure values using Sievert's law ($C_L = s \sqrt{p_{H_2}}$), using the solubility of pipeline steels: $s = 0.077 \, \text{wppm} \, \text{MPa}^{-0.5}$ \cite{Martin2020}. An excellent fit to the data is attained through the following choices of degradation law coefficients:  $\xi=0.12$, $\eta=9$, and $b=0.8$. The same hydrogen degradation law is applied to each of the weld regions, as no specific data is available, and this is likely to constitute a good first-order approximation.\\

Finally, an additional coupling is defined to capture how the hydrogen-containing environment is readily exposed to newly created crack surfaces as the crack grows. This is achieved following the work by D\'{\i}az \textit{et al.} \cite{diaz2025comsol}, where the hydrogen diffusivity is enhanced inside of crack regions, to simulate how the hydrogen gas will promptly occupy the space that has become available due to crack propagation. Accordingly, this enhanced diffusivity is defined as,
\begin{equation}\label{Dmov}
D_L^{\text{mov}} = D_L \left[ 1 + k_d \mathit{H}(\phi - \phi_{\text{th}}) \right] \, ,
\end{equation}
where $\mathit{H}()$ is the Heaviside function,  \(k_d \gg 1\) is the enhancement parameter, and \(\phi_{\text{th}} = 0.9\) is a threshold coefficient that controls the damage level above which it is assumed that hydrogen has progressed through the crack material.

\subsubsection{Numerical implementation}

The coupled deformation-diffusion-fracture formulation is implemented numerically in the finite element package \texttt{ABAQUS} using user subroutines. Specifically, and different to previous works, only integration point-level subroutines are used, enabling the use of \texttt{ABAQUS}'s in-built capabilities. The multi-trap hydrogen transport model is implemented by means of a \texttt{UMATHT} subroutine, using the code provided by Fernandez-Sousa \textit{et al.} \cite{fernandez2020analysis}. The elastic-plastic mechanical behaviour and its coupling with hydrogen were implemented via a \texttt{UMAT} subroutine. Finally, the evolution of the phase field variable was solved for with an additional \texttt{UMATHT} subroutine, exploiting the analogy with the heat transfer problem and adopting the \emph{twin-part} method recently presented by Navidtehrani \textit{et al.} \cite{NAVIDTEHRANI2025111363}. The deformation-fracture problem is solved in a monolithic way, while a multi-pass staggered scheme is adopted to couple it to the hydrogen transport problem. The reader is referred to Ref. \cite{NAVIDTEHRANI2025111363} for additional details on solution schemes and the \texttt{ABAQUS} implementation at the integration point level of coupled multi-field problems.

\subsection{Model validation} 
\label{sec:Validation}

\begin{figure}[H]
    \centering
    \includegraphics[width=0.93\textwidth]{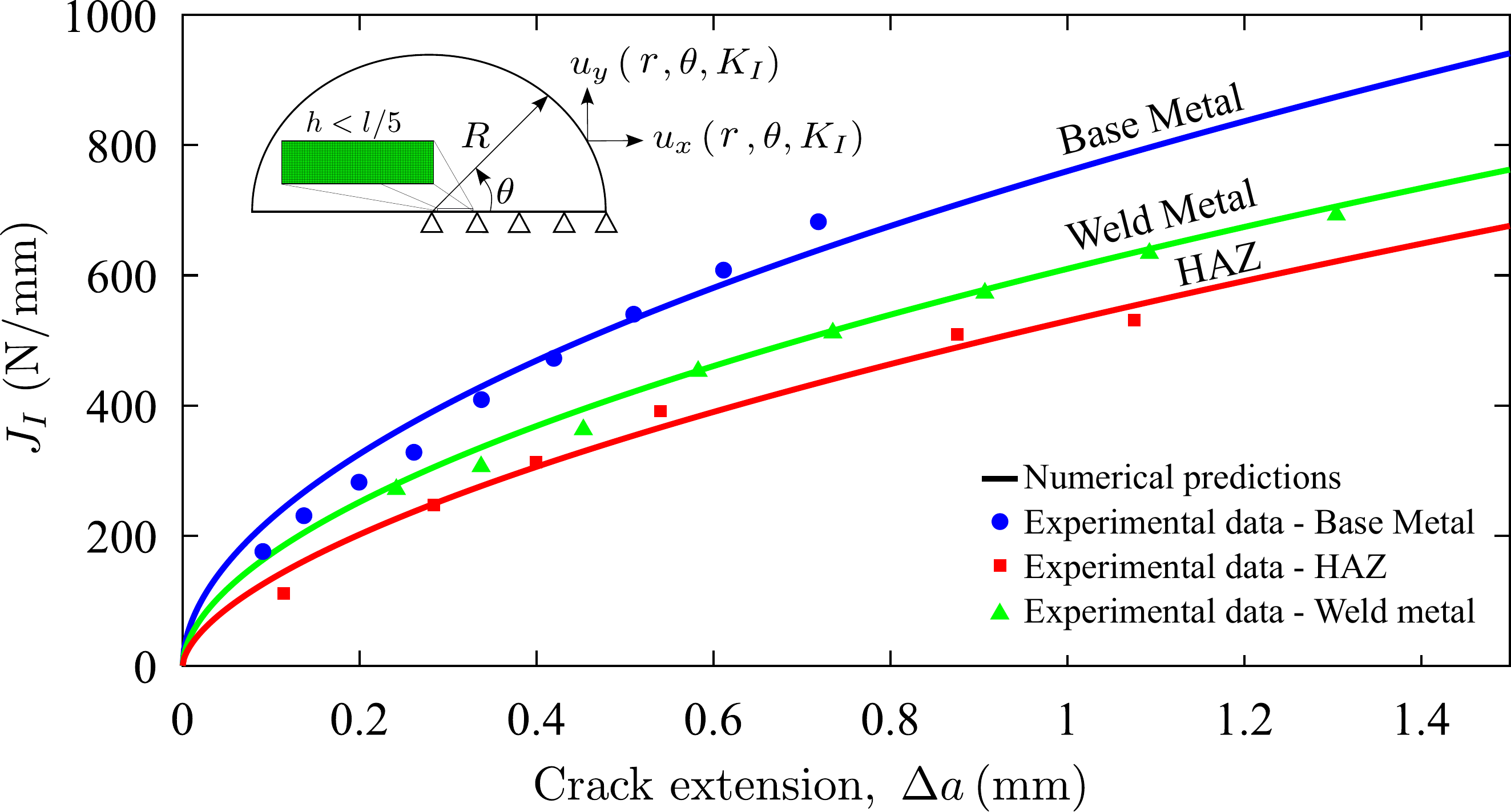}
    \caption{Computational predictions of crack growth resistance (J-R) curves for the three regions of the X80 pipeline steel weld: BM, WM, and HAZ. The numerical results, obtained with a boundary layer model, show a very good agreement with the experimental data from Ref. \cite{Yang2015}.}
    \label{fig:Rcurves}
\end{figure}

Our general framework is now particularised to the study of pipeline steels. In particular, we focus on X80 pipeline steel, for which sufficient information is available \cite{Yang2015}. The material properties for the BM, HAZ, and WM are provided in Table \ref{tab:material_properties}, as reported in the literature \cite{Yang2015}. While similar values are reported for the elastic and plastic properties, significant differences are observed in the values of the fracture toughness among the various weld regions, with the BM being the tougher region and the HAZ being the less fracture-resistant one (as expected). Poisson's ratio ($\nu$) is taken to be equal to 0.3 in all cases. It remains to define the value of the phase field length scale, which is determined by the choices of material strength $\sigma_c$ and toughness $G_c$, as per Eq. (\ref{eq:lc}). Following Ref. \cite{mandal2024computational}, the material strength is taken to be four times the material yield strength in the BM region ($\sigma_c = 4 \sigma_{y0}$). The $\sigma_c$ values for the WM and HAZ are then determined by noting that, upon assuming a similar degree of plastic dissipation with crack growth, a lower degree of strain hardening (as quantified by $N$) should result in a lower material strength \cite{Hutchinson1992}. With these considerations and based on the experimental data available (Table \ref{tab:material_properties} and Fig. \ref{fig:Rcurves}), the choices of $\sigma_c=3.55 \sigma_{y0}$ for the WM and $\sigma_c=3.75 \sigma_{y0}$ for the HAZ are made to accurately capture the degree of plastic dissipation with crack growth.

\begin{table}[H]
    \centering
    \begin{tabular}{|c|c|c|c|c|}
        \hline
        Region & \( E \) (MPa) & \( \sigma_{y0} \) (MPa) & \( N \) (-) & \( G_c \) (N/mm) \\ \hline
        Base Metal & 190480 & 570 & 0.10 & 90 \\ \hline
        HAZ         & 202010 & 598 & 0.08 & 50 \\ \hline
        Weld Metal  & 180300 & 688 & 0.07 & 57 \\ \hline
    \end{tabular}
    \caption{Mechanical and fracture properties adopted for the various weld regions of an X80 pipeline steel, taken from the experimental literature \cite{Yang2015}.}
    \label{tab:material_properties}
\end{table}

A boundary layer model is used to run virtual fracture experiments, upon the assumption of small-scale yielding conditions, and compare the outputs against the laboratory tests conducted in Ref. \cite{Yang2015}. Thus, a remote $K_I$ field (or, equivalently, $J_I$ field) is applied using William's elastic solution \cite{Williams1960}; i.e., the vertical and horizontal components of the displacement vector of the nodes located in the outer radius of the boundary layer are defined as,
\begin{equation}
u_x = K_I \frac{1 + \nu}{E} \sqrt{\frac{r}{2 \pi}} \left( (3 - 4\nu - \cos\theta) \cos\left(\frac{\theta}{2}\right) \right) \, ,
\end{equation}
\begin{equation}
u_y = K_I \frac{1 + \nu}{E} \sqrt{\frac{r}{2 \pi}} \left( (3 - 4\nu - \cos\theta) \sin\left(\frac{\theta}{2}\right) \right) \, ,
\end{equation}
where $r$ and $\theta$ are the coordinates of a polar coordinate system centered at the crack tip, and \( K_I \) is the mode I stress intensity factor, which in a plane strain solid is related to the \( J \)-integral by $J_I = K_I^2 (1 - \nu^2)/E$. To exploit the reflective symmetry of the boundary value problem with respect to the crack plane, only half of the full domain is simulated, ensuring that the outer radius is sufficiently large so as not to influence the results; i.e., \( R \gg R_p \), where \( R_p \) is Irwin's estimate of the plastic zone size,
\begin{equation}
R_p = \frac{1}{3\pi} \left( \frac{K_I}{\sigma_y} \right)^2 \, .
\end{equation}

In all simulations, the characteristic element size $h$ is chosen to be at least five times smaller than the phase field length scale $\ell$, to ensure mesh-independent results \cite{Kristensen2021}. Three models, with the same loading configuration and geometry but different material properties, are created to independently assess the ability of the numerical framework to predict the distinct fracture behaviour of the BM, HAZ, and WM regions. The crack extension ($\Delta a$) is measured as a function of the applied $J_I$, with the results being given in Fig. \ref{fig:Rcurves}, together with the experimental data.

Numerical predictions show a very good agreement with the experimental data obtained for each of the weld regions, demonstrating the ability of the model to accurately capture the initiation and growth of cracks in welded X80 pipeline steel. The results show that, by simulating elastic-plastic fracture in agreement with thermodynamic principles, the model naturally captures the increase in crack growth resistance observed with crack extension due to plastic dissipation, as a natural outcome of the model (i.e., without having to define \emph{a priori} how the toughness varies with crack extension).

\subsection{Predictions of girth weld integrity}
\label{Sec:ResultsFracture}

We proceed to employ the validated model to predict the structural integrity of welds in hydrogen transport pipelines under a wide range of relevant conditions. First, the boundary value problem considered is described in Section \ref{Sec:BVP}. Second, the behaviour of defect-free pipelines is assessed, to assess hydrogen transport and trapping, and determine reference critical failure pressures (Section \ref{Sec:DefectFree}). Finally, in Section \ref{Sec:DefectsResults}, a comprehensive study on the role of defects is conducted, where all possible defects (as allowed by the standards) are considered, spanning porosity, lack of penetration, imperfections, lack of fusion, root contraction, and undercutting defects. Their interplay with hydrogen is quantified, determining their impact in reducing admissible pressures in hydrogen transmission pipelines.

\subsubsection{Boundary value problem}
\label{Sec:BVP}

\begin{figure}[H]
    \centering
    \includegraphics[width=0.97\textwidth]{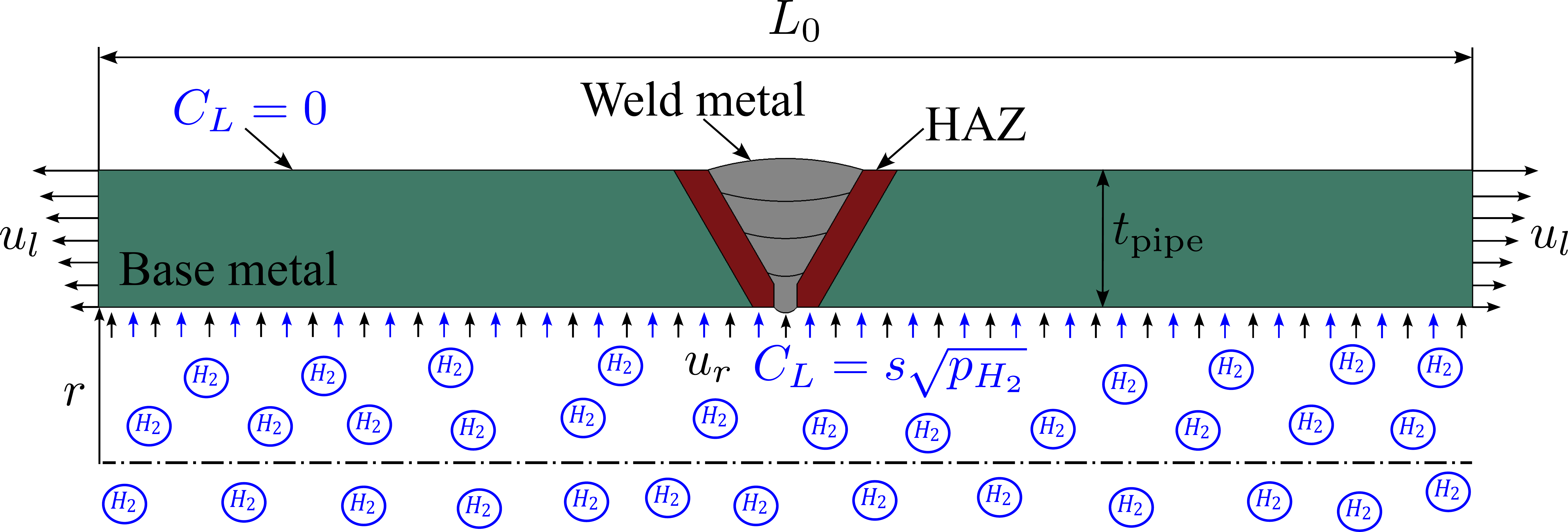}
    \caption{Boundary value problem mimicking the conditions of a girth weld in a hydrogen transport pipeline. The mechanical ($u_r$, $u_l$) and chemical ($C_L$) boundary conditions are illustrated, along with relevant dimensions: pipeline thickness $t_\text{pipe}$, inner radius $r_i$, and domain width $L_0$. The finite element model accounts for three weld regions with distinct material properties (see Table~\ref{tab:material_properties}).}
    \label{fig:BC}
\end{figure}

The integrity of girth welds in hydrogen pipelines is assessed through a deformation-diffusion-fracture model that uses the same dimensions as the weld process model (Section \ref{Sec:Thermo-Mechanical analysis}), enabling the coupling between the two. Thus, using a \texttt{SDVINI} subroutine, the residual stress state is transferred from the welding model to the weld integrity model, and the former also dictates the dimensions of the weld regions in the latter (as discussed in Section \ref{Sec:ResultsWeld}). As shown in Fig. \ref{fig:BC}, a 2D axisymmetric model is employed, which reproduces the conditions of girth welds in hydrogen pipelines. As such, a hydrogen concentration is defined in the interior of the pipeline, while hydrogen is expected to be able to exit the pipeline at its outer surface. The latter is captured by a $C_L=0$ boundary condition in the exterior boundary, while the hydrogen exposure boundary condition is given by Sievert's law, with the hydrogen concentration in the nodes in the interior surface being defined as $C_L=s \sqrt{p_{H_2}}$. The H$_2$ pressure ($p_{H_2}$) is slowly ramped up until failure (at a rate of 27 Pa/s), and the solubility is taken to be $s = 0.077 \, \text{wppm} \, \text{MPa}^{-0.5}$, as reported for pipeline steels \cite{Martin2020}. The mechanical boundary conditions capture, in a displacement-controlled setting, the loading resulting from the interior pressure. Integrating the Lame equations, the following relationships can be established between the pipeline pressure $p$ and the radial ($u_r$) and longitudinal ($u_l$) displacements:
\begin{equation}\label{ur}
    u_r = \frac{p r_i^2}{E t_{\text{pipe}} } \left( 1 - \frac{\nu}{2} \right) \, ,
\end{equation}
\begin{equation}\label{ul}
    u_l = \frac{p L_0 r_i}{2E t_{\text{pipe}}} \left( \frac{1}{2} - \nu \right) \, ,
\end{equation}
\noindent where $t_{\text{pipe}}$ is the pipeline thickness, $r_i$ the inner radius of the pipe, and $L_0$ is the width of the 2D axisymmetric model. These relationships are used to define $u_r$ and $u_l$ at the relevant surfaces (see Fig. \ref{fig:BC}), capturing in a 2D setting the 3D behaviour of a pipeline under an internal pressure $p$. As shown in Fig. \ref{fig:BC}, three different weld regions are considered, with distinct properties - those given in Table \ref{tab:material_properties}. The same finite element mesh is used for the welding and structural integrity simulations; i.e., the mesh is refined in the expected crack region, with the characteristic element length being 5 times smaller than the phase field length scale, to ensure mesh objective results.

\subsubsection{Critical hydrogen pressures for defect-free girth welds}
\label{Sec:DefectFree}

Let us begin the weld structural integrity analysis by considering the reference case of a defect-free pipeline. Two failure modes are considered: (i) fracture, as characterised by the propagation of a crack throughout the thickness of the pipeline, and (ii) plastic collapse, whereby the yielding pressure of the pipeline $p_{\text{yield}}=\sigma_y t_{\text{pipe}}/r_i$ is reached before fracture occurs. For the X80 pipeline steel considered here, $p_{\text{yield}}=30$ MPa, and thus calculations are stopped when this point is reached, ensuring that the above-defined relationships between the applied displacements and the pressure hold throughout the simulation.\\

\begin{figure}[H]
    \centering
    \includegraphics[width=1
    \textwidth]{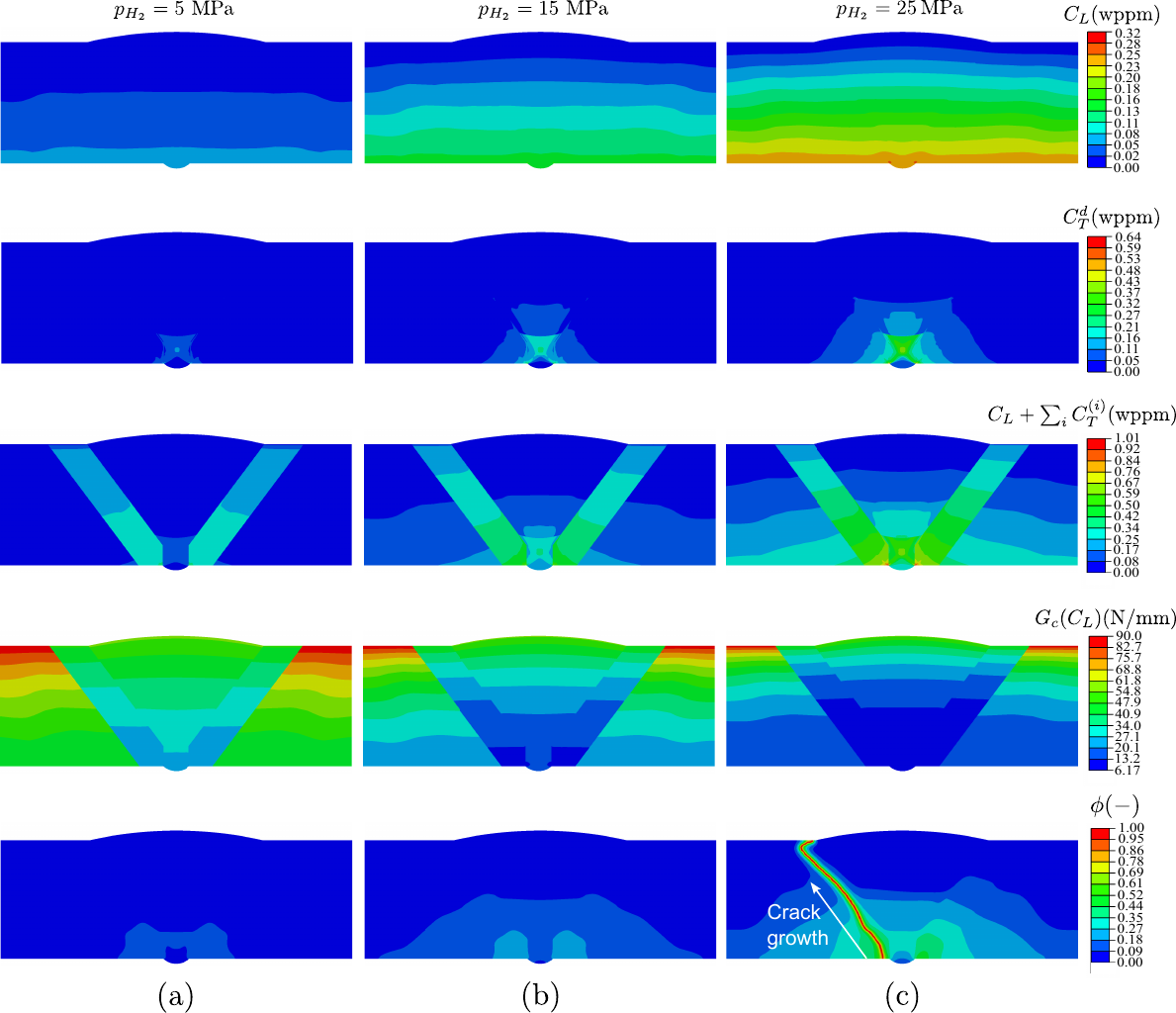}
    \caption{Contours of relevant properties for an initially defect-free pipeline under three different H$_2$ pressures (times): 5 MPa (left column), 15 MPa (centre column) and 25 MPa (right column). The contours show the spatial and temporal distribution of the lattice hydrogen concentration $C_L$ (first row), the hydrogen concentration trapped at dislocations $C_T^{(d)}$ (second row), the total hydrogen concentration (trapped and lattice) $C_L + \sum_i C_T^{(i)}$ (third row), the hydrogen-degraded material toughness $G_c (C_L)$ (fourth row), and the phase field fracture order parameter $\phi$ (fifth row). Failure occurs at a critical pressure of $p_f=25$ MPa, as indicated by the growth of a crack ($\phi=1$) from the weld root to the outer surface.}
    \label{fig:FractureRS}
\end{figure}

Representative results are shown in Fig. \ref{fig:FractureRS} in the form of contours of relevant variables: the lattice hydrogen concentration ($C_L$), the hydrogen trapped at dislocations ($C_T^{(d)}$), the total hydrogen concentration (lattice and trapped; $C_L + \sum_i C_T^{(i)}$), the hydrogen-dependent toughness ($G_c(C_L)$), and the phase field fracture variable ($\phi$). The phase field contours (fifth row, Fig. \ref{fig:FractureRS}) reveal that a crack develops near the weld root and grows along the HAZ until reaching the outer surface of the pipeline. Hydrogen-assisted failure occurs at a pressure of  25 MPa, below the yield pressure of 30 MPa. This is in contrast with the simulations conducted in the absence of hydrogen ($C_L=0 \,\, \forall \, \textbf{x}$), where the pipeline fails by plastic collapse. This earlier fracture in a hydrogen-containing environment is due to the interplay between the (lattice) hydrogen content and the material toughness. As shown in the first row of Fig. \ref{fig:FractureRS}, $C_L$ rises in the inner surface with increasing pressure, as per Sievert's law, and then diffuses throughout the pipeline thickness. No significant differences are observed in the lattice hydrogen concentration distribution along the weld regions. This is in agreement with expectations, given the similar apparent diffusivities of the BM, WM and HAZ. The resulting $C_L$ distribution renders a heterogeneous distribution of the material toughness $G_c (C_L)$ (see the fourth row in Fig. \ref{fig:FractureRS}), which facilitates cracking near the inner surface and in the HAZ region. The regions of higher susceptibility (lower $G_c (C_L)$) are dictated by both the hydrogen distribution and the inherent properties of each weld region. The results emphasise the lower hydrogen-assisted fracture resistance of the HAZ region, as recently quantified experimentally by Chalfoun \textit{et al.} \cite{Chalfoun2025}. The interplay between hydrogen and traps can also be visualised. Consider first the $C_T^{(d)}$ contours, presented in the second row of Fig. \ref{fig:FractureRS}. The dislocation trap density and the hydrogen trapped at dislocations are highest near the weld root, where the highest tensile residual stresses are attained after the weld process (see Fig. \ref{fig:WeldingResults}c). Comparison of the first three rows of Fig. \ref{fig:FractureRS} reveals that the hydrogen trapped in dislocations accounts for a very significant percentage of the trapped hydrogen, which is much higher than the lattice one. This indicates that welding residual stresses dominate hydrogen trapping in welds. The total (lattice and trapped) hydrogen concentration contours (third row in Fig. \ref{fig:FractureRS}) also reveal significant variations across the different regions of the weld, reflecting the interplay between microstructural heterogeneities and trapping.

\subsubsection{Resolving the interplay between pre-existing defects and critical hydrogen pressures}
\label{Sec:DefectsResults}

Defects are inherently present in welded joints, yet their role in compromising the structural integrity of welded components is difficult to predict. The deleterious effect of welding defects on structural integrity is exacerbated in the presence of hydrogen, as brittle failure becomes more likely (vs plastic collapse), and the presence of defects (stress concentrators) will in turn increase the local concentrations of trapped and lattice hydrogen. Typical defects resulting from the welding process include pores, lack of penetration defects, imperfections, lack of fusion defects, root contraction, and undercutting. Due to their quasi-inevitable nature, these are allowed in the standards; e.g., standards such as API 1104 \cite{API1041} and ISO 5817:2023 \cite{Iso5} provide guidelines for acceptable levels of welding imperfections and specify the maximum allowable flaw sizes. Hence, defects within the range classified as allowable in the standards are highly likely to be present in the natural gas pipelines being considered for hydrogen transport. To evaluate, for the first time, the impact that these defects can have under the more demanding conditions of hydrogen transport, we proceed to assess each of them, quantifying the critical pressure at which failure can happen and comparing the outcome with the result in the absence of hydrogen and the defect-free condition (Section \ref{Sec:DefectFree}).\\

First, let us consider the influence that porosity can have. The presence of small pores is a natural outcome of the welding process, typically as a result of gas trapping during solidification. This can be exacerbated due to improper gas shielding, contamination, or incorrect welding parameters. To span a wide range of relevant scenarios, calculations are conducted for the boundary value problem presented in Section \ref{Sec:BVP}, considering different degrees of porosity (pore volume fractions), with pores randomly distributed throughout the domain. A pore diameter of approximately 5 \textmu m is considered, consistent with welding practice. In the model, this is achieved by assigning an initial value of $\phi=1$ to a given percentage of integration points within the weld metal (e.g., to 1\% of the integration points, for $f_p=1$\%). Three scenarios are considered: no pores (i.e., the reference case assessed in Section \ref{Sec:DefectFree}), 0.5\% volume fraction of pores ($f_p=0.5$\%), and 1\% volume fraction of pores ($f_p=1$\%). These choices span realistic scenarios as the standards allow for up to 1\% porosity in welds \cite{Iso5}. 

    \begin{figure}[H]
    \centering
    \includegraphics[width=0.6\textwidth]{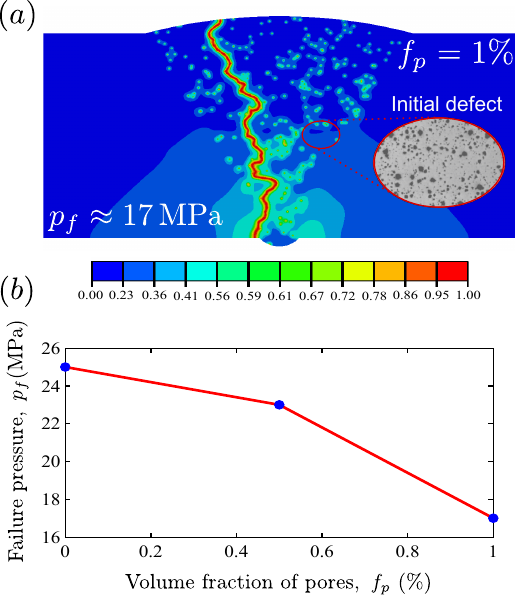}
    \caption{Impact of porosity on the structural integrity of hydrogen transport pipelines. The results are provided as: (a) phase field $\phi$ contours for the representative case of 1\% porosity, showing an inset of a microscopy image emphasising typical porosity levels found in welds, and  (b) estimations of failure pressure versus the pore volume fraction $f_p$ (in \%). Porosity levels that are considered acceptable for natural gas pipelines appear to bring a very significant reduction of load carrying capacity in H$_2$ environments.}

    \label{fig:Porosity}
\end{figure}

The results obtained are shown in Fig. \ref{fig:Porosity}, where the phase field contours are provided for the representative case of $f_{p} = 1\%$ (Fig. \ref{fig:Porosity}a) and all the results obtained are complied in a failure pressure vs pore volume fraction plot (Fig. \ref{fig:Porosity}b). Porosity is found to have a very significant effect on structural integrity, localising hydrogen concentration, and facilitating fracture. Specifically, the critical pressure is found to be as low as 17 MPa for the case of 1\% porosity, a scenario allowed in the standards \cite{Iso5}. This is a load carrying capacity reduction of approximately 32\%. The crack trajectory is also significantly influenced by porosity, with the crack no longer growing through the HAZ (as in the reference case, see Section \ref{Sec:DefectFree}) but instead following a higher porosity path through the WM, from its initiation point at the weld root. \\

\begin{figure}[H]
    \centering
    \includegraphics[width=1
    \textwidth]{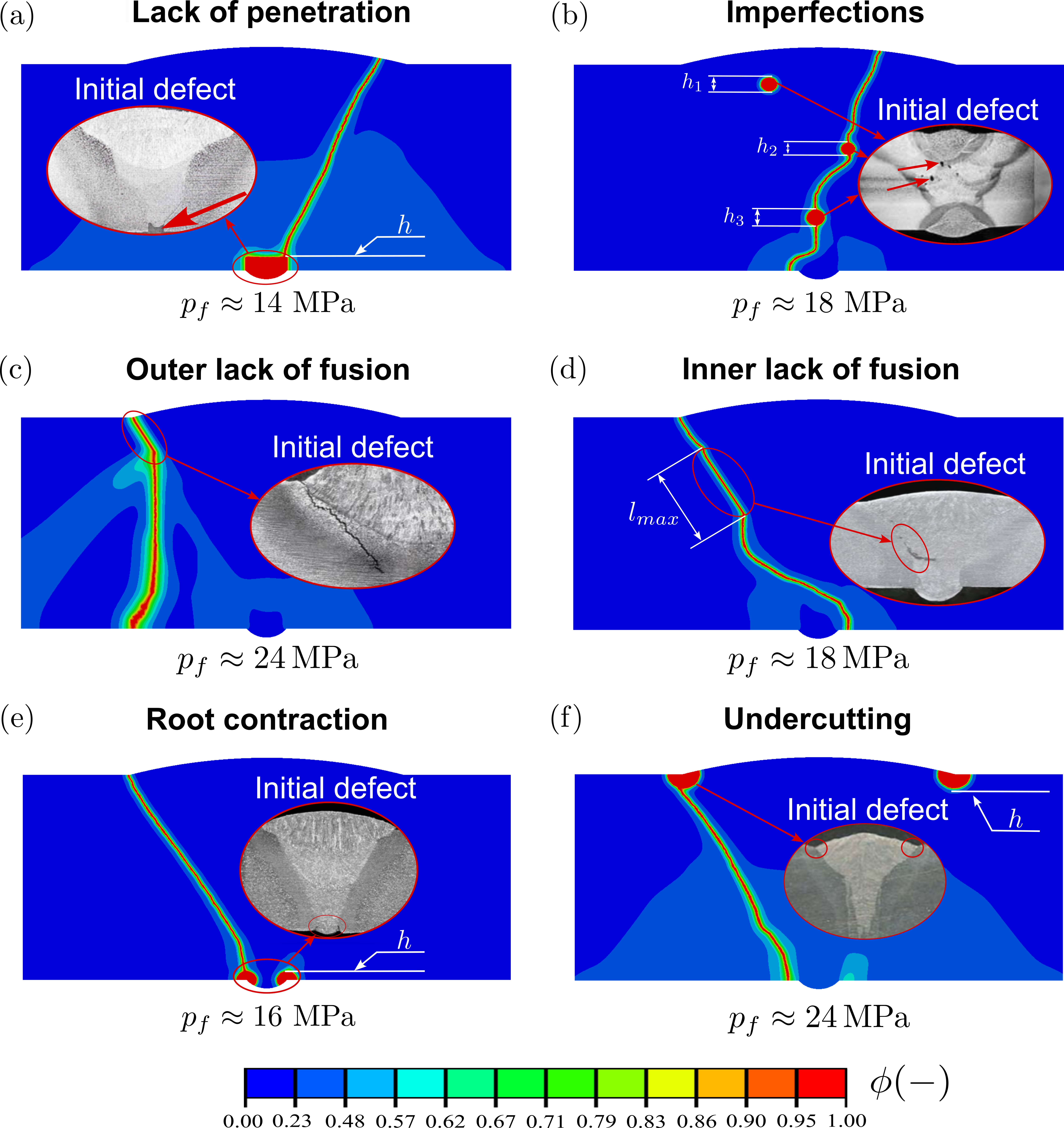}
    \caption{Resolving the interplay between weld defects and hydrogen-assisted pipeline failures. Cracking patterns, as described by phase field contours, and failure pressures ($p_f$) for: (a) lack of penetration defects, (b) weld imperfections, (c) outer lack of fusion defects, (d) inner lack of fusion defects, (e) root contraction defects, and (f) undercutting. In each case, a micrograph is included to showcase the visual appearance of each defect type. Defects considered acceptable in natural gas operation significantly reduce the maximum allowable pressure in H$_2$.}
    \label{fig:Defects}
\end{figure}

Next, let us address other defects that are commonly present in pipeline welds, under the conditions (defect dimensions) that are allowed in the standards \cite{API1041,Iso5}. The numerical predictions of the model are given in Fig. \ref{fig:Defects}, in terms of phase field contours (cracking trajectories) and failure pressures ($p_f$) for the following welding defects: lack of penetration, imperfections, lack of fusion (outer and inner), contraction at the root, and undercutting. Each of these scenarios is discussed below.\\

\noindent \textbf{Lack of penetration} happens when the WM does not extend through the entire thickness of the joint. It is commonly exacerbated by insufficient heat input, incorrect electrode angle, or too high welding speed. As per the standards \cite{Iso5}, the depth of lack of penetration defects cannot exceed $h=2$ mm. Hence, simulations are conducted with a 2 mm lack of penetration defect as initial condition, as shown in Fig. \ref{fig:Defects}a. The results show that the crack initiates at the edges of the weld defect and grows along the WM, with failure occurring at a pressure of 14 MPa. This entails a reduction of load carrying capacity of $\sim$44\%, relative to the reference, defect-free case. \\

\noindent \textbf{Weld imperfections} include various types of minor defects such as slag inclusions and surface irregularities. The presence of individual defects is accepted if the sum of their transverse area does not exceed 20\% of the pipeline thickness: $\sum_{i} h_{i} \leq 0.2 t_{\text{pipe}}$. Hence, for this pipe, which has a thickness of $t_{\text{pipe}} = 12$ mm, we consider three defects, randomly distributed within the WM region, with diameters $h_1 = 0.8$ mm, $h_2 = 0.7$ mm, and $h_3 = 0.9$ mm, such that they are within the maximum permissible values of the standard; the total projected defect length is $2.4$ mm, which corresponds to 20\% of the pipe thickness. The results, shown in Fig. \ref{fig:Defects}b, reveal a cracking path that is governed by these imperfections, with the failure pressure being $p_f=18$ MPa, a roughly 28\% reduction relative to the defect-free case.\\

\noindent \textbf{Lack of fusion} defects arise when WM fails to properly adhere to the BM or preceding weld passes. This typically occurs due to suboptimal heat input, poor welding technique (e.g., improper electrode angle or travel speed), or inadequate joint preparation (such as the presence of oxides or surface contaminants). Two scenarios are assessed here, to cover the most typical cases: an external lack of fusion defect (Fig. \ref{fig:Defects}c) and an internal one (Fig. \ref{fig:Defects}d). A defect length of 4 mm, corresponding to the maximum allowable value in the standard \cite{Iso5}, is considered only for the internal pass, while in the external case, the defect size is limited to the thickness of the weld bead of the last pass. These lack of fusion defects change the crack initiation location from the weld root to the defect tip. It can also be seen that internal lack of fusion defects are more detrimental than those at the outer surface, with the latter resulting in failure at a pressure of$p_f=18$ MPa, while the former can withstand pressures of 24 MPa. This implies a 28\% and 4\% reduction in load-carrying capacity, relative to the defect-free case, for the outer and inner lack of fusion defects, respectively.\\ 

\noindent \textbf{Root contraction} occurs due to the shrinkage of the WM as it cools and solidifies. This defect can lead to the formation of gaps or voids at the root of the weld, with the standards allowing root contraction defects of up to 0.5 mm in diameter \cite{Iso5}. The results obtained considering two root contraction defects are given in Fig. \ref{fig:Defects}e. As expected, the crack initiates from one of the two root contraction defects, and grows mainly along the HAZ. The pipeline weld fails at a H$_2$ pressure of 16 MPa, which is a 36\% reduction relative to the defect-free case.\\ 

\noindent \textbf{Undercutting} defects are typically characterised by a groove that forms at the weld edges, where the BM is melted away but not filled by the WM. As shown in Fig. \ref{fig:Defects}f, outer surface undercutting can result in stress concentrators that lead to cracking. The API standard \cite{API1041} is the most restrictive regarding undercutting defects, specifying that their length must not exceed 0.8 mm. The results show that cracks initiate at the undercutting defect and propagate through the HAZ. This leads to a reduction of approximately $4\%$ in the failure pressure, with 
$p_f$ being 24 MPa. This is the highest critical pressure attained across all the defects examined, suggesting that undercutting defects are the least harmful ones in hydrogen transport pipelines.\\

\begin{figure}[H]
    \centering
    \includegraphics[width=0.7\textwidth]{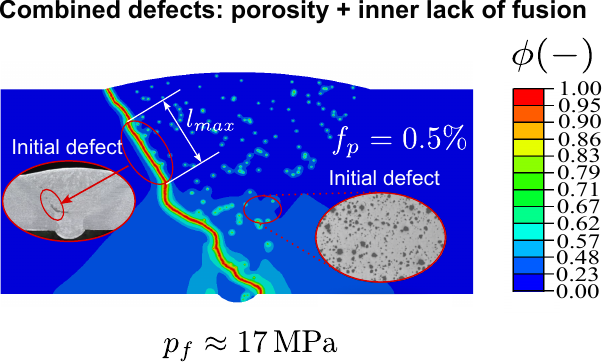}
    \caption{ Synergistic effect between inner lack of fusion ($l_\mathrm{max} = 4$ mm) and porosity (0.5\%) defects. The phase field contours show the cracking pattern resulting from the interaction of both defect types, revealing an intensified crack propagation that leads to a 39\% reduction in the failure pressure ($p_f = 17$ MPa), relative to the defect-free case, yet only a 5\% drop relative to the lack of fusion case. The results indicate that the coexistence of defects further promotes hydrogen-assisted cracking, yet the overall behaviour is still mainly governed by the most harmful defect.}
    \label{fig:CombinedDefect}
\end{figure}

The results obtained show that lack of penetration and root contraction defects are the most harmful ones, while undercutting and outer lack of fusion defects lead to the maximum allowable hydrogen transport pressures. Finally, we also examine a \emph{combined defects} case, to assess the synergistic effect of multiple defects coexisting within the welded joint. As illustrated in Fig. \ref{fig:CombinedDefect}, a case study combining internal lack of fusion and porosity defects was considered and taken as representative. Individually, these defects exhibit failure pressures of 18 MPa and 23 MPa, respectively; however, when both are present simultaneously, the failure pressure drops to $p_f=17$ MPa. This corresponds to an overall reduction of $\sim39\%$ relative to the defect-free configuration. These results highlight that the coexistence of different defect types further compromises the structural integrity of welded joints, as their combined influence leads to a more severe degradation in load-carrying capacity than the individual contribution of each defect. However, the drop in failure pressure is not very significant as the overall behaviour is mainly governed by the most harmful defect.

\section{Conclusions}
\label{Sec:Conclusions}

We have presented a novel computational framework combining thermo-mechanical weld process modelling with coupled deformation-diffusion-fracture simulations. The framework considers elastic-plastic deformation and fracture, using the phase field method, as well as multi-trapping phenomena, enabling to resolve the interplay between residual stresses and hydrogen trapping for the first time. The computational framework presented is particularised to the study of girth welds in X80 steel pipelines aimed at hydrogen transport. First, the weld process model is run to determine the residual stress state and the dimensions of the various weld regions: base metal (BM), weld metal (WM), and heat-affected zone (HAZ). The results capture the expected trends in thermal and mechanical fields. Then, structural integrity simulations are conducted to quantify the detrimental effect of hydrogen in welds. Key findings include:
\begin{itemize}
    \item Residual stresses dominate hydrogen trapping in welds. The hydrogen trapped in dislocations is found to be the primary contributor to the trapped hydrogen concentration, and its maximum values are attained near the weld root, where the welding residual stresses are highest. 
    
    \item Elastic-plastic phase field fracture modelling can predict the crack growth resistance behaviour of the various weld regions (BM, WM, HAZ), accurately capturing the role of plastic dissipation in increasing toughness with crack growth. 
    
    \item The combination of hydrogen, residual stresses, and heterogeneity of weld properties brings in a change in failure model, from plastic collapse to rapid fracture, even in the absence of initial defects.

    \item The microstructural heterogeneity of the weld regions results in distinct hydrogen trapping and embrittlement susceptibility, with cracking localising along the HAZ.   

    \item Porosity was found to have a significant effect, with allowed porosity levels (1\%) reducing by 32\% the maximum H$_2$ pipeline pressure (vs the defect-free case). Porosity also impacts the crack trajectory, shifting the highest susceptibility from the HAZ to the regions with the highest porosity.

    \item  Weld defects that are considered non-critical in natural gas pipelines, and thus allowed by the standards, can significantly compromise the structural integrity of hydrogen pipelines. Specifically, relative to the defect-free case, lack of penetration defects can bring down the failure pressure by 44\%, root contraction defects by 36\%,weld imperfections by 28\%, inner lack of fusion defects by 28\%, outer lack of fusion defects by 4\%, and undercutting by 4\%. Cracking patterns are also found to be very sensitive to these defects, governing crack trajectories and initiation sites.

    \item The presence of multiple co-existent defect types further compromises the structural integrity of hydrogen transport pipelines, although the synergistic effect is not significant, as the behaviour is mainly governed by the most harmful defect.

\end{itemize}   

Overall, the results obtained emphasise the need for stringent quality control in welds of hydrogen transport pipelines and a change in standards and protocols. The insight gained and the framework developed can be instrumental in mapping safe regimes of operation in hydrogen transmission pipelines, given the challenges associated with mimicking real working conditions (residual stresses, non-conventional defects) in laboratory settings. Potential avenues of future work include the consideration of external defects (such as dents) and degradation due to cyclic loading (hydrogen-assisted fatigue). Process optimisation can also be carried out through advanced weld process modelling, capable of predicting the emergence of defects that can be critical in hydrogen environments.

\section*{Acknowledgments}

\noindent The authors acknowledge funding from the Regional Government of Asturias (grant SV-PA-21-AYUD/2021/50985), from the Ministry of Science, Innovation, and Universities of Spain (grant MCINN-23-CPP2021-008986) and from the Ministry of Economy and Competitiveness of Spain (project PID2021-124768OB-C22). L. Castro additionally acknowledges financial support from the IUTA (grant SV-21-GIJÓN-1-18). E.\ Mart\'{\i}nez-Pa\~neda additionally acknowledges financial support from UKRI's Future Leaders Fellowship programme [grant MR/V024124/1], from the National Research Foundation of Korea (NRF) through the MSIT grant RS-2024-00397400, and from the UKRI Horizon Europe Guarantee programme (ERC Starting Grant \textit{ResistHfracture}, EP/Y037219/1).

\section*{CRediT authorship contribution statement}
\begin{justify}
\textbf{L. Castro:} Conceptualization, Investigation, Methodology, Software, Validation, Formal analysis, Writing – original draft, Writing – review \& editing, Visualization. \textbf{Y. Navidtehrani:} Investigation, Methodology, Software, Writing – review \& editing. \textbf{C. Betegón:} Investigation, Funding acquisition, Supervision, Project administration, Writing – review \& editing. \textbf{E. Martínez-Pañeda:} Conceptualization, Investigation, Methodology, Resources, Supervision, Project administration, Funding acquisition, Writing – review \& editing.
\end{justify}

\section*{Declaration of Competing Interest}
\noindent The authors declare that they have no known competing financial interests or personal relationships that could have appeared to influence the work reported in this paper.


\begin{thebibliography}{10}
\expandafter\ifx\csname url\endcsname\relax
  \def\url#1{\texttt{#1}}\fi
\expandafter\ifx\csname urlprefix\endcsname\relax\def\urlprefix{URL }\fi
\expandafter\ifx\csname href\endcsname\relax
  \def\href#1#2{#2} \def\path#1{#1}\fi

\bibitem{van2020hydrogen}
S.~Van~Renssen, The hydrogen solution?, Nature Climate Change 10~(9) (2020)
  799--801.

\bibitem{yu2024hydrogen}
H.~Yu, A.~D{\'\i}az, X.~Lu, B.~Sun, Y.~Ding, M.~Koyama, J.~He, X.~Zhou,
  A.~Oudriss, X.~Feaugas, et~al., Hydrogen embrittlement as a conspicuous
  material challenge- comprehensive review and future directions, Chemical
  Reviews 124~(10) (2024) 6271--6392.

\bibitem{Gangloff2003}
R.~P. Gangloff, Hydrogen-assisted cracking, in: I.~Milne, R.~Ritchie,
  B.~Karihaloo (Eds.), Comprehensive Structural Integrity Vol. 6, Elsevier
  Science, 2003, pp. 31--101.

\bibitem{harris2025hydrogen}
Z.~D. Harris, B.~P. Somerday, Hydrogen embrittlement of steels: Mechanical
  properties in gaseous hydrogen, International Materials Reviews (2025)
  09506608251338698.

\bibitem{chen2024hydrogen}
Y.-S. Chen, C.~Huang, P.-Y. Liu, H.-W. Yen, R.~Niu, P.~Burr, K.~L. Moore,
  E.~Mart{\'\i}nez-Pa{\~n}eda, A.~Atrens, J.~M. Cairney, {Hydrogen trapping and
  embrittlement in metals-- A review}, International Journal of Hydrogen Energy
  147 (2025) 150033.

\bibitem{iwadate1977prediction}
T.~Iwadate, T.~Karaushi, J.~Watanabe, Prediction of fracture toughness of
  2$1/4$cr-1mo pressure steels from charpy v-notch test results, in: Flaw
  Growth and Fracture, ASTM International, 1977, pp. 493--506.

\bibitem{hosseini2017trapping}
Z.~S. Hosseini, M.~Dadfarnia, K.~A. Nibur, B.~P. Somerday, R.~P. Gangloff,
  P.~Sofronis, et~al., Trapping against hydrogen embrittlement, in: Proceedings
  of the 2016 International Hydrogen Conference: Materials Performance in
  Hydrogen Environments, ASME Press, New York, NY, 2017, pp. 71--80.

\bibitem{ronevich2015hydrogen}
J.~A. Ronevich, Hydrogen embrittlement of pipeline steels in base metal and
  welds., Tech. rep., Sandia National Lab.(SNL-CA), Livermore, CA (United
  States) (2015).

\bibitem{diaz2021influence}
A.~D{\'\i}az, I.~Cuesta, C.~Rodr{\'\i}guez, J.~Alegre, Influence of
  non-homogeneous microstructure on hydrogen diffusion and trapping simulations
  near a crack tip in a welded joint, Theoretical and Applied Fracture
  Mechanics 112 (2021) 102879.

\bibitem{Chalfoun2025}
D.~R. Chalfoun, J.~Parker, M.~Gagliano, E.~Mart{\'\i}nez-Pa{\~n}eda, Tailored
  heat treatments to characterise the fracture resistance of critical weld
  regions in hydrogen transmission pipelines, (submitted).

\bibitem{alvaro2014hydrogen}
A.~Alvaro, V.~Olden, A.~Macadre, O.~M. Akselsen, Hydrogen embrittlement
  susceptibility of a weld simulated x70 heat affected zone under h2 pressure,
  Materials Science and Engineering: A 597 (2014) 29--36.

\bibitem{Ronevich2021}
J.~A. Ronevich, E.~J. Song, B.~P. Somerday, C.~W. {San Marchi},
  {Hydrogen-assisted fracture resistance of pipeline welds in gaseous
  hydrogen}, International Journal of Hydrogen Energy 46~(10) (2021)
  7601--7614.

\bibitem{ronevich2018fatigue}
J.~A. Ronevich, C.~R. D'Elia, M.~R. Hill, Fatigue crack growth rates of x100
  steel welds in high pressure hydrogen gas considering residual stress
  effects, Engineering Fracture Mechanics 194 (2018) 42--51.

\bibitem{tekkaya2024multi}
B.~Tekkaya, M.~D{\"o}lz, S.~M{\"u}nstermann, Multi-scale approach to hydrogen
  susceptibility based on pipe-forming deformation history, International
  Journal of Mechanical Sciences 282 (2024) 109625.

\bibitem{Bortot2024InvestigationEnvironment}
P.~Bortot, M.~Ortolani, M.~Sileo, E.~Escorza, M.~Connolly, Z.~N. Buck,
  A.~Chandra, {Investigation of the fracture resistance of high strength
  ferritic steel welds in gaseous hydrogen environment}, International Journal
  of Hydrogen Energy 136 (2025) 777--788.

\bibitem{zhou2024multiscale}
H.~Zhou, M.~Kinefuchi, Y.~Takashima, K.~Shibanuma, Multiscale modelling
  strategy for predicting fatigue performance of welded joints, International
  Journal of Mechanical Sciences 284 (2024) 109751.

\bibitem{cui2025fracture}
C.~Cui, L.~Cupertino-Malheiros, Z.~Xiong, E.~Mart{\'\i}nez-Pa{\~n}eda, On the
  fracture mechanics validity of small scale tests, Engineering Fracture
  Mechanics (2025) 111321.

\bibitem{seo2022fracture}
K.-W. Seo, J.-H. Hwang, Y.-J. Kim, K.-S. Kim, P.-S. Lam, Fracture toughness
  prediction of hydrogen-embrittled materials using small punch test data in
  hydrogen, International Journal of Mechanical Sciences 225 (2022) 107371.

\bibitem{kim2024fracture}
J.-Y. Kim, K.-W. Seo, Y.-J. Kim, K.-S. Kim, Fracture analysis of
  hydrogen-embrittled api x52 pipes at low temperature, International Journal
  of Mechanical Sciences 276 (2024) 109374.

\bibitem{Serebrinsky2004}
S.~Serebrinsky, E.~A. Carter, M.~Ortiz, {A quantum-mechanically informed
  continuum model of hydrogen embrittlement}, Journal of the Mechanics and
  Physics of Solids 52~(10) (2004) 2403--2430.

\bibitem{yu2016uniform}
H.~Yu, J.~S. Olsen, A.~Alvaro, V.~Olden, J.~He, Z.~Zhang, A uniform hydrogen
  degradation law for high strength steels, Engineering Fracture Mechanics 157
  (2016) 56--71.

\bibitem{DelBusto2017}
S.~del Busto, C.~Beteg{\'{o}}n, E.~Mart{\'{i}}nez-Pa{\~{n}}eda, {A cohesive
  zone framework for environmentally assisted fatigue}, Engineering Fracture
  Mechanics 185 (2017) 210--226.

\bibitem{Martinez-Paneda2018}
E.~Mart{\'{i}}nez-Pa{\~{n}}eda, A.~Golahmar, C.~F. Niordson, A phase field
  formulation for hydrogen assisted cracking, Computer Methods in Applied
  Mechanics and Engineering 342 (2018) 742--761.

\bibitem{diddige2025phase}
V.~Diddige, S.~Roth, B.~Kiefer, Phase-field modeling of hydrogen-promoted
  fracture: Natural incorporation of hydrostatic stress dependencies via a
  chemical potential-based variational formulation, Computer Methods in Applied
  Mechanics and Engineering 445 (2025) 118143.

\bibitem{tan2025phase}
Y.~Tan, F.~Peng, P.~Li, C.~Liu, J.~Zhao, X.~Li, A phase-field fracture model
  for piezoelectrics in hydrogen-rich environment, International Journal of
  Mechanical Sciences 291 (2025) 110092.

\bibitem{Kristensen2021}
P.~K. Kristensen, C.~F. Niordson, E.~Mart{\'{i}}nez-Pa{\~{n}}eda, {An
  assessment of phase field fracture: Crack initiation and growth},
  Philosophical Transactions of the Royal Society A: Mathematical, Physical and
  Engineering Sciences 379~(2203) (2021).

\bibitem{yin2025diffusive}
Y.~Yin, H.~Yu, H.~Yan, S.~Zhu, Diffusive-length-scale adjustable phase field
  fracture model for large/small structures, International Journal of
  Mechanical Sciences 285 (2025) 109839.

\bibitem{kristensen2020phase}
P.~K. Kristensen, E.~Mart{\'\i}nez-Pa{\~n}eda, Phase field fracture modelling
  using quasi-newton methods and a new adaptive step scheme, Theoretical and
  Applied Fracture Mechanics 107 (2020) 102446.

\bibitem{mandal2021fracture}
T.~K. Mandal, V.~P. Nguyen, J.-Y. Wu, C.~Nguyen-Thanh, A.~de~Vaucorbeil,
  Fracture of thermo-elastic solids: Phase-field modeling and new results with
  an efficient monolithic solver, Computer Methods in Applied Mechanics and
  Engineering 376 (2021) 113648.

\bibitem{kristensen2020applications}
P.~K. Kristensen, C.~F. Niordson, E.~Mart{\'\i}nez-Pa{\~n}eda, Applications of
  phase field fracture in modelling hydrogen assisted failures, Theoretical and
  Applied Fracture Mechanics 110 (2020) 102837.

\bibitem{hai2024novel}
L.~Hai, H.~Zhang, P.~Wriggers, Y.-j. Huang, Y.~Feng, P.~Junker, A novel
  semi-explicit numerical algorithm for efficient 3d phase field modelling of
  quasi-brittle fracture, Computer Methods in Applied Mechanics and Engineering
  432 (2024) 117416.

\bibitem{yi2024phase}
D.~Yi, Z.~Yang, L.~Yi, J.~Liu, C.~Yang, L.~Gou, N.~Zheng, X.~Li, D.~Fu,
  Z.~Huang, Phase-field model of hydraulic fracturing in thermoelastic--plastic
  media, International Journal of Mechanical Sciences 283 (2024) 109750.

\bibitem{zhao2024phase}
J.~Zhao, Y.~F. Cheng, A phase field method for predicting hydrogen-induced
  cracking on pipelines, International Journal of Mechanical Sciences 283
  (2024) 109651.

\bibitem{mandal2024computational}
T.~K. Mandal, J.~Parker, M.~Gagliano, E.~Mart{\'\i}nez-Pa{\~n}eda,
  Computational predictions of weld structural integrity in hydrogen transport
  pipelines, International Journal of Hydrogen Energy 136 (2025) 923--937.

\bibitem{joboxford}
J.~Wijnen, J.~Parker, M.~Gagliano, E.~Martínez-Pañeda, A computational
  framework to predict weld integrity and microstructural heterogeneity:
  Application to hydrogen transmission, Materials \& Design 249 (2025) 113533.

\bibitem{wijnen2025virtual}
J.~Wijnen, J.~Parker, M.~Gagliano, E.~Mart{\'\i}nez-Pa{\~n}eda, Virtual failure
  assessment diagrams for hydrogen transmission pipelines, International
  Journal of Hydrogen Energy 149 (2025) 149984.

\bibitem{Xu2024}
T.~Xu, S.~Guo, B.~Fu, X.~Ma, H.~Han, Y.~Li, H.~Jiang, {Hydrogen diffusion
  simulation of the X80 pipeline steel girth weld zone considering the
  synergistic effect of the structure–stress–concentration field},
  Engineering Failure Analysis 160 (2024) 108--205.

\bibitem{Huang2024}
Z.~Huang, J.~Li, L.~Wang, L.~Lei, X.~Huang, Z.~Yin, {Welding residual stress
  analysis of the X80 pipeline: Simulation and validation}, Mechanical Sciences
  15 (2024) 305--314.

\bibitem{Yang2015}
Y.~Yang, L.~Shi, Z.~Xu, H.~Lu, X.~Chen, X.~Wang, {Fracture toughness of the
  materials in welded joint of X80 pipeline steel}, Engineering Fracture
  Mechanics 148 (2015) 337--349.

\bibitem{AmericanPetroleumInstitute2018}
{American Petroleum Institute}, {Line Pipe API Specification 5L - Forty-Sixth
  Edition}~(April) (2018) 205.

\bibitem{Mathias2013}
L.~L. Mathias, D.~F. Sarzosa, C.~Ruggieri, {Effects of specimen geometry and
  loading mode on crack growth resistance curves of a high-strength pipeline
  girth weld}, International Journal of Pressure Vessels and Piping 111-112
  (2013) 106--119.

\bibitem{Gaudet2015}
M.~J. Gaudet, The tensile properties and toughness of microstructures relevant
  to the haz of x80 linepipe steel girth welds, Phd thesis, University of
  Alberta, Edmonton, Canada (2015).

\bibitem{Yu2023}
D.~Yu, C.~Yang, Q.~Sun, L.~Dai, A.~Wang, H.~Xuan, {Impact of process parameters
  on temperature and residual stress distribution of X80 pipe girth welds},
  International Journal of Pressure Vessels and Piping 203 (2023) 104939.

\bibitem{Seles}
K.~Sele{\v{s}}, M.~Peri{\'{c}}, Z.~Tonkovi{\'{c}}, {Numerical simulation of a
  welding process using a prescribed temperature approach}, Journal of
  Constructional Steel Research 145 (2018) 49--57.

\bibitem{Parmar2016}
C.~Parmar, C.~Gill, B.~Pellereau, P.~Hurrell, {Simulation of a Multi-Pass
  Dissimilar Metal Nozzle to Pipe Weld Using Abaqus 2D Weld GUI and Comparison
  with Measurements Chandrahas}, in: Proceedings of the ASME 2016 Pressure
  Vessels and Piping Conference (PVP2016), ASME, Vancouver, British Columbia,
  Canada, 2016.

\bibitem{Garcin2016}
T.~Garcin, M.~Militzer, W.~J. Poole, L.~Collins, {Microstructure model for the
  heat-affected zone of X80 linepipe steel}, Materials Science and Technology
  (United Kingdom) 32~(7) (2016) 708--721.

\bibitem{isfandbod2021mechanism}
M.~Isfandbod, E.~Mart{\'\i}nez-Pa{\~n}eda, A mechanism-based multi-trap phase
  field model for hydrogen assisted fracture, International Journal of
  Plasticity 144 (2021) 103044.

\bibitem{cupertino2024suitability}
L.~Cupertino-Malheiros, T.~K. Mandal, F.~Th{\'e}bault,
  E.~Mart{\'\i}nez-Pa{\~n}eda, On the suitability of single-edge notch tension
  (sent) testing for assessing hydrogen-assisted cracking susceptibility,
  Engineering Failure Analysis 162 (2024) 108360.

\bibitem{krom1999hydrogen}
A.~H. Krom, R.~W. Koers, A.~Bakker, Hydrogen transport near a blunting crack
  tip, Journal of the Mechanics and Physics of Solids 47~(4) (1999) 971--992.

\bibitem{diaz2024explicit}
A.~D{\'\i}az, J.~Alegre, I.~Cuesta, Z.~Zhang, Explicit implementation of
  hydrogen transport in metals, International Journal of Mechanical Sciences
  273 (2024) 109195.

\bibitem{cupertino2023hydrogen}
L.~Cupertino-Malheiros, A.~Oudriss, F.~Th{\'e}bault, M.~Piette, X.~Feaugas,
  Hydrogen diffusion and trapping in low-alloy tempered martensitic steels,
  Metallurgical and Materials Transactions A 54~(4) (2023) 1159--1173.

\bibitem{TaylorQuinney}
G.~I. Taylor, H.~Quinney, The latent energy remaining in a metal after cold
  working, Proceedings of the Royal Society of London. Series A, Containing
  Papers of a Mathematical and Physical Character 143~(849) (1934) 307--326.

\bibitem{R.A.Oriani1974}
R.~Oriani, P.~Josephic, Equilibrium aspects of hydrogen-induced cracking of
  steels, Acta Metallurgica 22 (1974) 1065--1074.

\bibitem{garcia2024tds}
E.~Garc{\'\i}a-Mac{\'\i}as, Z.~D. Harris, E.~Mart{\'\i}nez-Pa{\~n}eda, Tds
  simulator: a matlab app to model temperature-programmed hydrogen desorption,
  International Journal of Hydrogen Energy 94 (2024) 510--524.

\bibitem{Zheng2023}
S.~Zheng, Y.~Qin, W.~Li, F.~Huang, Y.~Qiang, S.~Yang, L.~Wen, Y.~Jin, Effect of
  hydrogen traps on hydrogen permeation in x80 pipeline steel a joint
  experimental and modelling study, International Journal of Hydrogen Energy
  48~(12) (2023) 4773--4788.

\bibitem{Sun2021}
Y.~Sun, Y.~F. Cheng, {Hydrogen permeation and distribution at a high-strength
  X80 steel weld under stressing conditions and the implication on pipeline
  failure}, International Journal of Hydrogen Energy 46~(44) (2021)
  23100--23112.

\bibitem{diaz2016coupled}
A.~D{\'\i}az, J.~Alegre, I.~Cuesta, Coupled hydrogen diffusion simulation using
  a heat transfer analogy, International Journal of Mechanical Sciences 115
  (2016) 360--369.

\bibitem{jemblie2017review}
L.~Jemblie, V.~Olden, O.~M. Akselsen, A review of cohesive zone modelling as an
  approach for numerically assessing hydrogen embrittlement of steel
  structures, Philosophical Transactions of the Royal Society A: Mathematical,
  Physical and Engineering Sciences 375~(2098) (2017) 20160411.

\bibitem{fernandez2020analysis}
R.~Fern{\'a}ndez-Sousa, C.~Beteg{\'o}n, E.~Mart{\'\i}nez-Pa{\~n}eda, Analysis
  of the influence of microstructural traps on hydrogen assisted fatigue, Acta
  Materialia 199 (2020) 253--263.

\bibitem{LUFRANO1998827}
J.~Lufrano, P.~Sofronis, D.~Symons, Hydrogen transport and large strain
  elastoplasticity near a notch in alloy x-750, Engineering Fracture Mechanics
  59~(6) (1998) 827--845.

\bibitem{martinez2016strain}
E.~Mart{\'\i}nez-Pa{\~n}eda, C.~F. Niordson, R.~P. Gangloff, Strain gradient
  plasticity-based modeling of hydrogen environment assisted cracking, Acta
  Materialia 117 (2016) 321--332.

\bibitem{Gilman1969}
{Gilman JJ.}, {Micromechanics of flow in solids}, McGraw-Hill, 1969.

\bibitem{dadf2011}
M.~Dadfarnia, P.~Sofronis, T.~Neeraj, Hydrogen interaction with multiple traps:
  Can it be used to mitigate embrittlement?, International Journal of Hydrogen
  Energy 36~(16) (2011) 10141--10148, european Fuel Cell 2009.

\bibitem{Shang2024}
J.~Shang, J.~Guo, Z.~Hua, B.~Xing, T.~Cui, H.~Wei, Effects of plastic
  deformation on hydrogen trapping and hydrogen distribution in x80 pipeline
  steel, International Journal of Hydrogen Energy 136 (2025) 1306--1316.

\bibitem{Li2020}
X.~Li, X.~Ma, J.~Zhang, E.~Akiyama, Y.~Wang, X.~Song, {Review of Hydrogen
  Embrittlement in Metals: Hydrogen Diffusion, Hydrogen Characterization,
  Hydrogen Embrittlement Mechanism and Prevention}, Acta Metallurgica Sinica
  (English Letters) 33~(6) (2020) 759--773.

\bibitem{H.K.D.H.Bhadeshia2016}
{H.K.D.H. Bhadeshia}, {Prevention of Hydrogen Embrittlement in Steels}, ISIJ
  International 56 (2016) 24--36.

\bibitem{ZHANG2024111764}
P.~Zhang, M.~Laleh, A.~E. Hughes, R.~K. Marceau, T.~Hilditch, M.~Y. Tan, Effect
  of microstructure on hydrogen embrittlement and hydrogen-induced cracking
  behaviour of a high-strength pipeline steel weldment, Corrosion Science 227
  (2024) 111764.

\bibitem{Yan2014}
C.-Y. Yan, C.-Y. Liu, G.-Y. Zhang, Simulation of hydrogen diffusion in welded
  joint of x80 pipeline steel, Journal of Central South University 21~(12)
  (2014) 4432--4437.

\bibitem{bourdin2000numerical}
B.~Bourdin, G.~A. Francfort, J.-J. Marigo, Numerical experiments in revisited
  brittle fracture, Journal of the Mechanics and Physics of Solids 48~(4)
  (2000) 797--826.

\bibitem{tanne2018crack}
E.~Tann{\'e}, T.~Li, B.~Bourdin, J.-J. Marigo, C.~Maurini, Crack nucleation in
  variational phase-field models of brittle fracture, Journal of the Mechanics
  and Physics of Solids 110 (2018) 80--99.

\bibitem{molnar2020toughness}
G.~Moln{\'a}r, A.~Doitrand, R.~Estevez, A.~Gravouil, Toughness or strength?
  regularization in phase-field fracture explained by the coupled criterion,
  Theoretical and applied fracture mechanics 109 (2020) 102736.

\bibitem{Amor2009}
H.~Amor, J.~J. Marigo, C.~Maurini, {Regularized formulation of the variational
  brittle fracture with unilateral contact: Numerical experiments}, Journal of
  the Mechanics and Physics of Solids 57~(8) (2009) 1209--1229.

\bibitem{Marchi2011}
C.~S. Marchi, B.~P. Somerday, K.~A. Nibur, D.~G. Stalheim, T.~Boggess,
  S.~Jansto, {Fracture resistance and fatigue crack growth of X80 pipeline
  steel in gaseous hydrogen}, American Society of Mechanical Engineers,
  Pressure Vessels and Piping Division (Publication) PVP 6~(PARTS A AND B)
  (2011) 841--849.

\bibitem{Marchi2022}
C.~S. Marchi, J.~Ronevich, {Implications of Gaseous Hydrogen on Welded
  Construction of Pipelines SAND2022-0820 PE}~(January) (2022).

\bibitem{Shang2021}
J.~Shang, J.~Z. Wang, W.~F. Chen, H.~T. Wei, J.~Y. Zheng, Z.~L. Hua, L.~Zhang,
  C.~H. Gu, {Different effects of pure hydrogen vs. hydrogen/natural gas
  mixture on fracture toughness degradation of two carbon steels}, Materials
  Letters 296 (2021) 88--91.

\bibitem{Martin2020}
M.~L. Martin, M.~J. Connolly, F.~W. DelRio, A.~J. Slifka, {Hydrogen
  embrittlement in ferritic steels}, Applied Physics Reviews 7~(4) (2020)
  041301.

\bibitem{diaz2025comsol}
A.~D{\'\i}az, J.~M. Alegre, I.~I. Cuesta, E.~Mart{\'\i}nez-Pa{\~n}eda, A comsol
  framework for predicting hydrogen embrittlement, part ii: Phase field
  fracture, Engineering Fracture Mechanics 319 (2025) 111008.

\bibitem{NAVIDTEHRANI2025111363}
Y.~Navidtehrani, C.~Betegón, E.~Martínez-Pañeda, A generalised framework for
  phase field-based modelling of coupled problems: Application to
  thermo-mechanical fracture, hydraulic fracture, hydrogen embrittlement and
  corrosion, Engineering Fracture Mechanics 326 (2025) 111363.

\bibitem{Hutchinson1992}
J.~W. Hutchinson, V.~Tvergaard, {The relation between crack growth resistance
  and fracture process parameters elastic-plastic solids}, Journal of the
  Mechanics and Physics of Solids 40~(6) (1992) 1377--1397.

\bibitem{Williams1960}
M.~L. Williams, {The bending stress distribution at the base of a stationary
  crack}, Journal of Applied Mechanics, Transactions ASME 28~(1) (1960) 78--82.

\bibitem{API1041}
American Petroleum Institute, API Std 1104,Welding of Pipelines and Related
  Facilities, 22nd Edition (2023).

\bibitem{Iso5}
European Standard, Welding. Fusion-welded joints in steel, nickel, titanium and
  their alloys (beam welding excluded). Quality levels for imperfections (ISO
  5817:2023). (2023).

\end{thebibliography}



\end{document}